\documentclass{LMCS}

\def\doi{8 (2:07) 2012}
\lmcsheading%
{\doi}
{1--21}
{}
{}
{Mar.~21, 2012}
{Jun.~\phantom01, 2012}
{}
\usepackage{verbatim}
\usepackage{graphicx}
\usepackage{epstopdf}
\usepackage{bussproofs}
\usepackage{color} 
\usepackage{dsfont}
\usepackage{stmaryrd}
\usepackage{amsthm}
\usepackage{mathtools}
\usepackage{amssymb}
\usepackage{hyperref}
\usepackage{pstricks,pst-node,pst-tree}
\usepackage{enumerate}
\usepackage[all]{xy}

\newtheorem{teorema}{Teorema}[section]

\newtheorem{theorem}[teorema]{Theorem}
\newtheorem{lemma}[teorema]{Lemma}
\newtheorem{proposition}[teorema]{Proposition}

\theoremstyle{definition} 
\newtheorem{definition}[teorema]{Definition}

\theoremstyle{remark} 
\newtheorem{remark}[teorema]{Remark}

\newcommand{\bydef}{ \stackrel{\mathrm{def}}{=} }

\newcommand{\freccia}[1]{\stackrel{#1}{\longrightarrow}}

\newcommand{\diam}[1]{ \langle #1\rangle }
\newcommand{\quadrato}[1]{\left[ #1 \right]}

\newcommand{\arena}{\mathcal{A}}

\newcommand{\reward}{\textnormal{R}}
\newcommand{\priority}{\Pr}

\newcommand{\lfp}{\mathrm{lfp}}
\newcommand{\gfp}{\mathrm{gfp}}

\newcommand{\sem}[1] {  \llbracket #1 \rrbracket  }  
\newcommand{\gsem}[1]{\llparenthesis{\,#1\,}\rrparenthesis}

\newcommand\nodeC[1]{*+[o][F]{#1}}

\newcommand\addLabelUL[1]{\ar@{}[]+UR|(1){~\makebox[0pt][l]{$\mathbf{#1}$}}}

\newcommand\addLabelUR[1]{\ar@{}[]+UR|(1){~\makebox[0pt][l]{$\mathbf{#1}$}}}

\newcommand\addLabelDL[1]{\ar@{}[]+UR|(1){~\makebox[0pt][l]{$\mathbf{#1}$}}}

\newcommand\addLabelDR[1]{\ar@{}[]+UR|(1){~\makebox[0pt][l]{$\mathbf{#1}$}}}

\newcommand\addDMD[2]{
	\ar@{-}[]+<#1pt,0pt>;[]+<0pt,#2pt>
	\ar@{-}[]+<0pt,#2pt>;[]+<-#1pt,0pt>
	\ar@{-}[]+<-#1pt,0pt>;[]+<0pt,-#2pt>
	\ar@{-}[]+<0pt,-#2pt>;[]+<#1pt,0pt>
}

\begin{document}

\title[On the Game Semantics of the the Probabilistic
  $\mu$-Calculus]{On the Equivalence of Game and Denotational
  Semantics for the Probabilistic $\mu$-Calculus\rsuper*}

\author{Matteo Mio} 
\address{LFCS, School of Informatics, University of Edinburgh}
\email{M.Mio@sms.ed.ac.uk}


\titlecomment{{\lsuper*}An extended abstract of this paper appeared in  \emph{Proceedings of the 7th Workshop on Fixed Points in Computer Science} (Luigi Santocanale ed.), 2010, pp. 83-87.}

\begin{abstract}
The \emph{probabilistic} (or \emph{quantitative}) modal $\mu$-calculus
is a fixed-point logic designed for expressing properties of probabilistic labeled transition systems (PLTS). Two semantics have been
studied for this logic, both assigning to every process state a value in
the interval $[0, 1]$ representing the probability that the
property expressed by the formula holds at the state. One semantics is \emph{denotational} and the other is a \emph{game semantics}, specified in terms of two-player stochastic games. The two semantics have been proved
to coincide on all finite PLTS's, but the  equivalence of the two semantics on arbitrary models has been open in literature.
In this  paper we prove that the equivalence indeed holds for arbitrary infinite models,
and thus our result strengthens the fruitful connection between denotational and game semantics. Our proof adapts the \emph{unraveling} or \emph{unfolding method}, a general proof technique for proving result of parity games by induction on their complexity.
\end{abstract}

\maketitle

\section{Introduction}
The modal $\mu$-calculus L${\mu}$ \cite{Kozen83} is a very expressive logic obtained by extending classical propositional modal logic with least and greatest fixed point operators. The logic L${\mu}$ has been extensively studied as it provides a very powerful tool for expressing properties of labeled transition systems \cite{Stirling96}. Encodings of many important temporal logic such as LTL, CTL and CTL${}^{*}$ into L${\mu}$ \cite{Dam94}, provided evidence for the very high expressive power of the calculus. A precise expressivity result was given in \cite{JW96}, where the authors showed that every formula of monadic second order logic over transition systems which does not distinguish between bisimilar models is equivalent to a formula of L${\mu}$. The logic L${\mu}$ has a simple denotational interpretation \cite{Stirling96}. However it is often  difficult to intuitively grasp the denotational meaning of a L${\mu}$ formula as the nesting of fixed point operators can induce very complicated properties. To alleviate this problem, another complementary semantics for the logic L${\mu}$, based on two-player (parity) games, has been studied in \cite{EJ91,Stirling96}. The two semantics have been proven to coincide and this allows us to pick the most convenient viewpoint when thinking about the logic L${\mu}$  \cite{EJ91,Stirling96}. 

In the last decade, a lot of research has focused on the study of reactive systems that exhibit some kind of probabilistic behavior, and logics for expressing their properties. Probabilistic labeled transition systems (PLTS's) \cite{S95} are a natural generalization of standard LTS's to the probabilistic scenario, as they allow both non-deterministic and (countable) probabilistic choices. A state $s$ in a PLTS can evolve by non-deterministically choosing one of the \emph{accessible} probability distributions $d$ (over process states) and then continuing its execution from the state $s^{\prime}$ with probability $d(s^{\prime})$. This combination of non-deterministic choices immediately followed by probabilistic ones, allows the modeling of concurrency, non-determinism and probabilistic behaviors in a natural way. PTLS's can be visualized using  graphs labeled with probabilities in a natural way \cite{HP2000,KNPV2009,Bartels02}. For example the PLTS depicted in Figure \ref{figura_intro_plts_1} models a system with two states $p$ and $q$. At the state $q$ no action can be performed. At the state $p$ the system can evolve non-deterministically either  to the state $q$ with probability $1$ (when the transition $p\freccia{a}d_{2}$ is chosen) or to the state $p$ with probability $\frac{1}{3}$ and to the state $q$ and with probability $\frac{2}{3}$ (when the transition $p\freccia{a}d_{1}$ is chosen). 
\begin{figure}[h!]
\begin{center}
$$
\SelectTips{cm}{}
	\xymatrix @=20pt {
		\nodeC{p} \ar@{->}[rr]^{a}  \ar@{->}[drr]^{a} & &  \nodeC{d_{1}} \ar@{.>}@/^10pt/[rr]^{\frac{2}{3}} \ar@{.>}@/_10pt/[ll]_{\frac{1}{3}} & &  \nodeC{q}  	\\
		& &   \nodeC{d_{2}} \ar@{.>}@/^0pt/[urr]^{1}\\
		}
$$
\end{center}
\caption{Example of PLTS}\label{figura_intro_plts_1}
\end{figure}

The probabilistic modal $\mu$-calculus pL${\mu}$, introduced in \cite{MM97,HM96,AM04}, is a generalization of L${\mu}$ designed for expressing properties of PLTS's. 
This logic was originally named the \emph{quantitative} $\mu$-calculus, but since other $\mu$-calculus-like logics, designed for expressing properties of non-probabilistic systems, have been given the same name \cite{FGK2010}, we adopt the \emph{probabilistic} adjective which is meant to emphasize that the models considered are PLTS's. The syntax of the logic pL$\mu$ coincides with that of the standard $\mu$-calculus. The denotational semantics  of pL${\mu}$ \cite{MM97,AM04} generalizes that of L$\mu$ by interpreting every formula $F$ as a map $\sem{F}\!:\!P\!\rightarrow\! [0,1]$, which assigns to each process state $p$ a \emph{degree of truth}. A key aspect of the denotational semantics of  \cite{MM97,AM04}  is the interpretation of conjunction, defined as ${\sem{F\wedge G}(p)}\!=\!{\min\{\sem{F}(p),\sem{G}(p)\}}$. This is not the only possible meaningful generalization of standard boolean conjunction to the real interval $[0,1]$. Indeed, different interpretations for the connectives of  pL${\mu}$ (including the one of \cite{MM97, AM04}) have been proposed in \cite{HM96}, and there is no \emph{a priori} good reason to prefer one in favour of the others.

In \cite{MM07}, the authors introduce an alternative semantics for the logic pL${\mu}$. This semantics, given in term of two player stochastic (parity) games \cite{zielonka04}, is a natural generalization of the two player (non stochastic) game semantics for the logic L$\mu$ \cite{Stirling96}. As in L$\mu$ games, the two players play a game starting from a
   configuration $\langle p, F\rangle$, where the objective for Player $1$ is
   to produce a path of configurations along which the outermost
   fixed point variable $X$ unfolded infinitely often is bound by a greatest
   fixed point in $F$. On a configuration of the form $\langle p,G_{1}\vee G_{2}\rangle$,
   Player $1$ chooses one of the disjuncts $G_{i}$, $i\!\in\!\{1,2\}$, by moving to the next
   configuration $\langle p, G_{i}\rangle$. On a configuration $\langle p,G_{1}\wedge G_{2}\rangle$,
   Player $2$ chooses a conjunct $G_{i}$ and moves to $\langle p, G_{i}\rangle$. On a configuration
   $\langle p, \mu X.G\rangle$ or  $\langle p, \nu X.G\rangle$ the game evolves to the configuration
   $\langle p, G\rangle$, after which, from any subsequent configuration $\langle q, X\rangle$ the game
   again evolves to $\langle q, G\rangle$. On configurations $\langle p, \diam{a}G\rangle$ and $\langle p, \quadrato{a}G\rangle$,
   Player $1$ and Player $2$ respectively choose a transition ${p}\!\freccia{a}\!{d}$ in
   the PLTS and move the game to $\langle d, G\rangle$. Here $d$ is a
   probability distribution over process-states (this is the key difference between
   pL$\mu$ and L$\mu$ games)  and the configuration $\langle d,G\rangle$ belongs to Nature, the probabilistic
   agent of the game, who moves on to the next configuration $\langle q,G\rangle$ with
   probability $d(q)$.  
This game semantics offers a clear operational interpretation for the \emph{properties} associated to the formulas, explained in terms of the interactions between the controller (Player $1$) and a hostile environment (Player $2$) in the context of the stochastic choices occurring in the PLTS (Nature).  The meaning of a pL$\mu$ formula $F$ at a state $p$ can be interpreted as expressing the (limit) probability for the controller to satisfy the \emph{property} specified by the formula. 

In \cite{MM07}, the equivalence of the denotational and game semantics of pL$\mu$  on all \emph{finite} models was proven. The proof, which adapts the standard technique of  \cite{Stirling96,EJ91} used to prove the equivalence of game and denotational semantics for L$\mu$, makes essential use of the fact that \emph{memoryless} and \emph{optimal} strategies exist in every \emph{finite} two-player stochastic game with \emph{parity} objectives \cite{zielonka04}. This property however, does not hold, in general, for two-player stochastic (parity) games of infinite size: optimal strategies may not exist, and an \emph{unbounded} amount of memory might be necessary even for playing $\epsilon$-optimally, i.e., for guaranteeing a probability of victory $\epsilon$-close to the optimal one. The general result, i.e., the equivalence of the game and denotational semantics of pL$\mu$ on arbitrary infinite models, is left open in \cite{MM07}.

In this  paper we prove that the equivalence indeed holds for arbitrary infinite models, thus strengthening the connection between denotational and game semantics. This result, given that  the pL$\mu$ games outlined above are natural generalization of standard L$\mu$ games, provides a justification for the denotational interpretation of the connectives of pL$\mu$ of  \cite{MM97,AM04}. The generalization of the result of \cite{MM07} to arbitrary infinite models is of practical interest since infinite state systems often provide natural abstractions for, e.g.,  infinite memory, infinite data-sets, \emph{etcetera}.  Our contribution consists in adapting a proof technique, called the \emph{unfolding method} \cite{Gradel2007,Santocanale2002}, which is adopted in \cite{FGK2010}  to prove a similar result for a $\mu$-calculus-like logic designed to express quantitative properties of (non probabilistic) labeled transition systems. While this is not a difficult adaptation, the result is worth noting since the question has been open in literature since \cite{MM07}. Moreover the differences between the games considered in \cite{FGK2010} and pL$\mu$ two-player stochastic games, e.g., the fact that \emph{Markov chains} are the outcomes of the games rather than just infinite paths, make this result not immediate from \cite{FGK2010}.

The rest of the paper is organized as follows. In Section \ref{basic}, we introduce some mathematical definitions. In Section \ref{syntax} we define the syntax and the denotational semantics of the logic pL${\mu}$ as in \cite{MM97,AM04}. In Section \ref{game_definitions}, we define the class of two-player stochastic parity games that are going to be used to give game semantics to the logic. In Section \ref{game_semantics}, we define the game semantics of pL${\mu}$ in terms of two-player stochastic parity games as in \cite{MM07,AM04} and state the main theorem which asserts the equivalence of the denotational and game semantics for pL$\mu$.  Lastly, in Section \ref{main_proof}, a detailed proof of the main theorem is given.
 
\section{Background definitions and notation}\label{basic}
\begin{definition}[Probability distributions] A (discrete) probability distribution $d$ over a set $X$ is a function $d\!:\!X\!\rightarrow\![0,1]$ such that $\sum_{x\in X}d(x)\!=\!1$. The \emph{support} of  $d$, denoted by $supp(d)$, is defined as the (necessarily countable)  set $\{x\!\in\! X\ | \ d(x)\!>\!0\}$. We denote with $\mathcal{D}(X)$ the set of probability distributions over $X$. We denote with $\delta_{x}$, for $x\!\in\! X$, the distribution over $X$ such that $supp(\delta_{x})\!=\!\{ x\}$, i.e., the unique distribution such that $\delta_{x}(x)=1$ and $\delta_{x}(y)=0$, for all $y\not=x$.
\end{definition}

\begin{definition}[PLTS \cite{S95}]\label{PLTS} Given a countable set $L$ of labels, a \emph{probabilistic labeled transition system} is a pair $\langle P, \{ \freccia{a}\}_{a\in L}\rangle$, where $P$ is a set (of arbitrary cardinality) of process-states, and ${\freccia{a}}\! \subseteq\! P\times \mathcal{D}(P)$ for every $a\!\in\! L$. As usual we write $p\freccia{a}d$ if ${(p,d)}\!\in\!{\freccia{a}}$. \end{definition}
The transition relation of a PLTS models the dynamics of the processes: $p\freccia{a}d$ means that the process $p$ can 
perform the atomic action $a\!\in\! L$ and then,  with probability $d (q)$, behave like the process $q$.  Probabilistic labeled transition system are a natural generalization of labeled transition systems to the probabilistic scenario: a standard LTS can be modeled as a PLTS in which every reachable distribution is of the form $\delta_{p}$, for some $p\!\in\! P$.

Given a set $X$, we denote with $2^{X}$ the set of all subsets $Y\!\subseteq\! X$. Given a complete lattice $(L,\leq)$, we denote with $\bigsqcup\!:2^{L}\!\rightarrow\! L$ and $\bigsqcap\!:2^{L}\!\rightarrow\! L$ the operations of join and meet respectively. A function $f\!:\!L\!\rightarrow\!L$ is \emph{monotone} if $x\!\leq\! y$ implies $f(x)\!\leq\! f(y)$, for every $x,y\!\in\!L$. The set of fixed points of any monotone function $f\!:\!L\rightarrow\!L$, ordered by $\leq$, is  a non-empty complete lattice \cite{Tarski1955}. We denote with $\lfp(f)$ and $\gfp(f)$ the least and the greatest fixed points of $f$, respectively. 

\begin{theorem}[Knaster-Tarski \cite{Tarski1955}]
Let $(L,\leq)$ be a complete lattice and $f\!:\!L\!\rightarrow\!L$ a monotone function. The following equalities hold:
\begin{enumerate}[(1)]
\item $\lfp(f) =\bigsqcup_{\alpha} f^{\alpha}$, where $f^{\alpha}=\bigsqcup_{\beta<\alpha}f(f^{\beta})$,
\item $\gfp(f) = \bigsqcap_{\alpha} f_{\alpha}$, where $f_{\alpha}=\bigsqcap_{\beta<\alpha}f(f_{\beta})$,
\end{enumerate}
where the greek letters $\alpha$ and $\beta$ ranges over ordinals.
\end{theorem}

In the following we assume standard notions of basic topology and basic measure theory which can be found in, e.g., \cite{Kechris,Tao_measuretheory}.

\section{The Probabilistic Modal $\mu$-Calculus}
\label{syntax}
Given a countable set $\textit{Var}$ of propositional variables ranged over by the letters $X, Y , Z$ and a set of labels $L$ ranged 
over by the letters $a, b, c$, the formulas of the logic pL$\mu$ (in positive form) are defined by the following 
grammar: 
\begin{center}
$F, G ::= X \ | \ \diam{a}F\ | \ \quadrato{a}F \ |  \ F\vee G\ | \ F\wedge G\ | \ \mu X.F\ | \ \nu X.F  $
\end{center}
As usual the operators  $\nu X.F$ and $\mu X.F$ bind the variable $X$ in $F$. A formula is \emph{closed} if it has no \emph{free} variables. 

\begin{definition}[Subformulas] We define the set $Sub(F)$ of \emph{subformualae} of $F$ by induction on the structure of $F$ as follows:
\begin{center}
\begin{tabular}{l l l l}
$\ \ \ $ & $Sub(   X  )$ &  $\bydef$ &  $\{  X   \}$\\ 
$\ \ \ $ & $Sub(  F_{1}\wedge F_{2}   )$ &  $ \bydef$ &  $ \{  F_{1}\wedge F_{2}   \} \cup Sub(F_{1} ) \cup Sub(F_{2})$\\
$\ \ \ $ & $Sub(  F_{1}\vee F_{2}   )$ &  $ \bydef$ &  $ \{  F_{1}\vee F_{2}   \} \cup Sub(F_{1} ) \cup Sub(F_{2})$\\
$\ \ \ $ & $Sub(   \quadrato{a}F_{1}  )$ &  $ \bydef$ &  $ \{ \quadrato{a}F_{1}    \} \cup Sub(F_{1})$\\
$\ \ \ $ & $Sub(   \diam{a}F_{1}  )$ &  $ \bydef$ &  $ \{ \diam{a}F_{1}    \} \cup Sub(F_{1})$\\
$\ \ \ $ & $Sub(  \nu X.F_{1}   )$ &  $ \bydef$ &  $ \{   \nu X.F_{1} \} \cup Sub(F_{1})$\\
$\ \ \ $ & $Sub(  \mu X.F_{1}   )$ &  $ \bydef$ &  $ \{   \mu X.F_{1} \} \cup Sub(F_{1})$
\end{tabular}
\end{center}
We say that $G$ is a subformula of $F$ if $G\!\in\! Sub(F)$.
\end{definition}
\begin{definition}[Normal Formula]
\label{normal_form}
We say that a formula $F$ is in \emph{normal form}, if every occurrence of a $\mu$ or $\nu$ binder binds a distinct variable, and no 
variable appears both free and bound. Every formula can be put in normal form by standard $\alpha$-renaming of the bound variables.
\end{definition}
For convenience we only consider, from now on,  formulas $F$ in normal form. This allows, for instance, the definition below to be given as follows:
\begin{definition}[Variable subsumption] Given a formula $F$, we say that $X$ \emph{subsumes} $Y$ in $F$, for $X\!\not =\! Y$, if $X$ and $Y$ are bound in $F$ by the sub-formulas $\star_{1} X.G$ and $\star_{2}  Y.H$ respectively, and $\star_{2} Y.H\in Sub(G)$, for $\star_{1},\star_{2}\in\{\mu,\nu\}$.

\end{definition}

Given a PLTS $\langle P, \{ \freccia{a} \}_{a\in L}\rangle$, we denote with $[0,1]^{P}$ or $P\rightarrow[0,1]$  the complete lattice of functions from $P$  to the real interval $[0, 1]$ with the pointwise order. A function $\rho\!:\! \textit{Var}\!\rightarrow\! [0,1]^{P}$ is called a $[0,1]$-valued \emph{interpretation}, or just an \emph{interpretation}, of the variables. Given a function $f\!:\! P\!\rightarrow\! [0,1]$ we denote with $\rho[f /X]$ the interpretation that assigns $f$ to the variable $X$, and $\rho (Y)$ to all other variables $Y$.

\begin{definition}[\cite{MM07}]\label{denotationa_sem_def}
The denotational semantics $\sem{F}_{\rho}\!:\! P\! \rightarrow\! [0,1]$ of the pL$\mu$ formula $F$ under the interpretation $\rho$,  is defined by structural induction on $F$ as follows:
\begin{center}
\begin{tabular}{l l l}
$\sem{X}_{\rho}(p)$ & $=$ & $\rho(X)(p)$\\
$\sem{G \vee H}_{\rho}(p)$ & $=$  & $\sem{ G}_{\rho}(p) \sqcup \sem{H}_{\rho}(p)$\\
$\sem{G \wedge H}_{\rho}(p)$ & $=$  & $\sem{ G}_{\rho}(p) \sqcap \sem{H}_{\rho}(p)$\\
$\sem{\diam{a}G}_{\rho}(p) $ & $=$ & $  \displaystyle \bigsqcup_{p\freccia{a}d} \big( \sum_{q\in supp(d)} d(q)\! \cdot\! \sem{G}_{\rho}(q) \big)$\\
$\sem{\quadrato{a}G}_{\rho}(p) $ & $=$ & $  \displaystyle \bigsqcap_{p\freccia{a}d} \big( \sum_{q\in supp(d)} d(q)\! \cdot\! \sem{G}_{\rho}(q) \big)$\\
$\sem{\mu X. G}_{\rho}(p) $ & $=$ & $  \lfp\big(\lambda f. ( \sem{G}_{\rho[ f/X]})\big)(p)$ \\
$\sem{\nu X. G}_{\rho}(p) $ & $=$ & $  \gfp\big( \lambda f. ( \sem{G}_{\rho[ f/X]})\big)(p)$
\end{tabular}
\end{center}
It is easy to verify that the interpretation assigned to every pL$\mu$ operator is monotone. Thus, the existence of the least and greatest fixed points is guaranteed by the Knaster-Tarski theorem. 
\end{definition}

The main novelty of \cite{MM07,HM96} in the definition of the semantics of pL$\mu$ resides in the interpretation of the modalities $\diam{a}$ and $\quadrato{a}$, for $a\!\in\! L$. The definitions resemble the corresponding ones for L$\mu$ (see, e.g., \cite{Stirling96}) but, crucially, in PLTS's transitions lead to probability distributions over processes, rather than processes. The most natural way to interpret the meaning of a formula $G$ at a probability distribution $d$ is to consider the \emph{expected probability} of the formula $G$ holding at a process $q$, associated by the random choice over processes induced with $d$, and this is formalized by the weighted sums in the definition above.

\begin{remark}
As it is common practice  when dealing with fixed point logics such as the modal $\mu$-calculus, we presented the syntax of pL$\mu$ in \emph{positive form}, i.e., without including a negation operator. This simplifies the presentation of the denotational semantics because all formulas in positive form are interpreted as monotone functions. A negation operator on (closed) pL$\mu$ formulas can be defined by induction on the structure of the formula, by exploiting the dualities between the connectives of the logic, in such a way that $\sem{\neg F}_{\rho}(p)= 1-\sem{F}_{\rho}(p)$, for all formulas $F$ and process states $p$. We omit the routine details.
\end{remark}

\section{Two Player Stochastic Parity Games}
\label{game_definitions}
In this section we introduce the class two-player stochastic games used to give game semantics to the logic pL$\mu$. This material in standard, and follows similar presentations, as in, e.g., \cite{zielonka04}.

A two-player turn-based stochastic game (or just a $2\frac{1}{2}$-player game) is played on some \emph{arena} ${\arena}\!=\!{\langle (S,E), (S_{1},S_{2},S_{N} ),\pi\rangle}$ where $(S,E)$ is a directed graph with (possibly uncountable) set of states $S$ and transition relation $E\!\subseteq\! S\times S$. The sets $S_{1}$, $S_{2}$, $S_{N}$ form a partition of $S$ and  $\pi\!:\!S_{N}\!\rightarrow\!\mathcal{D}(S)$ is called the probabilistic transition function. For every state $s\!\in\! S$, we denote with $E(s)$ the (possibly infinite) set  $\{ s^{\prime} \ | \ (s,s^{\prime})\!\in\! E \}$ of successors of $s$. We require that for all $s\!\in\! S_{N}$, the equality $E(s)\!=\!supp(\pi(s))$ holds. This implies that the set of successors of a state $s\!\in\! S_{N}$ is non-empty and at most countable. We denote with $S_{t}$ the set of \emph{terminal states}, i.e., those $s\!\in\! S$ such that $E(s)\!=\!\emptyset$. 

The game is played on the arena $\arena$ by three players named Player $1$, Player $2$ and Nature, the probabilistic agent of the game. The states in $S_{1}$ are under the control of Player $1$, the states in $S_{2}$ are under the control of Player $2$, and the states in $S_{N}$ are probabilistic, i.e., under the control of   Nature. At a state $s \!\in\! S_{1}$, if $s\not \in S_{t}$, Player $1$ chooses  a successor from the set $E(s)$; if $s\!\in\! S_{t}$ the game ends. Similarly,   at a state $s \!\in\! S_{2}$, if $s\not \in S_{t}$, Player $2$ chooses  a successor from the set $E(s)$; if $s\!\in\! S_{t}$ the game ends. At a state $s\!\in\! S_{N}$, a successor state is  probabilistically chosen according with the probability distribution $\pi(s)$. The outcome of a \emph{play} of the three players is a path in $\arena$, either countably-infinite or finite (ending in a terminal state), which we call a \emph{completed} path.

\begin{definition}
We denote with $\mathcal{P}^{\omega}$ and $\mathcal{P}^{<\omega}$ the sets of infinite and finite (non empty) paths  in $\arena$. Given a finite path $\vec{s}\!\in\!\mathcal{P}^{<\omega}$ we denote with $\mathit{last}(\vec{s})$ the last state $s\!\in\! S$ of  $\vec{s}$. We write $\vec{s}\lhd \vec{t}$, with $\vec{s},\vec{t}\!\in\!\mathcal{P}^{<\omega}$, if $\vec{t}\!=\!\vec{s}.s$, for some $s\!\in\! S$, where  the \emph{dot} symbol denotes the concatenation operator. We denote with $\mathcal{P}^{t}$ the set of finite paths ending in a terminal state, i.e., the set of paths $\vec{s}$ such that $\mathit{last}(\vec{s})\!\in\! S_{t}$.  We denote with $\mathcal{P}_{1}^{<\omega}$, $\mathcal{P}_{2}^{<\omega}$ and $\mathcal{P}_{N}^{<\omega}$ the sets of finite paths $\vec{s}$ such that $last(\vec{s})\!\in \!S_{1}$, $last(\vec{s})\!\in\! S_{2}$ and $\mathit{last}(\vec{s})\!\in\! S_{N}$ respectively. We denote with $\mathcal{P}$ the set $\mathcal{P}^{\omega}\cup \mathcal{P}^{t}$ and we refer to this set as the set of \emph{completed} paths in $\arena$. Given a finite path $\vec{s}\!\in\! \mathcal{P}^{<\omega}$, we denote with $O_{\vec{s}}$ the set of all completed paths having $\vec{s}$ as prefix. We consider the standard topology on $\mathcal{P}$ where the  basis for the open sets is given by the clopen sets $O_{\vec{s}}$, for $\vec{s}\!\in\! \mathcal{P}^{<\omega}$. This is a $0$-dimensional space and, if $S$ is countable, it is a Polish space.
We denote with $(\mathcal{P},\Omega)$ the Borel $\sigma$-algebra induced by the topology on $\mathcal{P}$, i.e., the smallest $\sigma$-algebra on $\mathcal{P}$ containing all the open sets.
\end{definition}

To specify the reward assigned to Player $1$ when a given completed path $\vec{s}$ is the outcome of a play, we introduce the notion of payoff function.

\begin{definition}
A \emph{(Borel) payoff function} for the arena $\arena$ is a Borel-measurable function $\Phi\!:\!\mathcal{P}\!\rightarrow\! [0,1]$. 
\end{definition}
The value $\Phi(\vec{s})$, for a given $\vec{s}\!\in\!\mathcal{P}$, should be understood as the reward assigned to Player $1$  when $\vec{s}$ is the outcome of a play in $\arena$.

\begin{definition}[Two player stochastic game] A \emph{two-player turn-based stochastic game} (or just a $2\frac{1}{2}$-player game) is a pair $\langle \arena,\Phi\rangle$, where $\Phi$ is a payoff function for the arena $\arena$.
\end{definition}
The goal of Player $1$ in the game $\langle \arena,\Phi\rangle$ is to maximize their payoff, while the \emph{dual} goal of Player $2$ is to minimize the payoff assigned to Player $1$.

When working with stochastic games, it is useful  to look at the possible outcomes of a play up to the behavior of Nature. This is done by introducing  the notion of Markov chain in $\arena$, whose precise formulation is given by the following definitions.

\begin{definition}[Tree in $\arena$]
A \emph{tree} in the arena $\arena$ is a collection $T\subseteq \mathcal{P}^{<\omega}$ of finite paths in $\arena$, such that
\begin{enumerate}[(1)]
\item $T$ is down-closed: if $\vec{s}\!\in\! T$ and $\vec{t}$ is a prefix of $\vec{s}$,  then $\vec{t}\!\in\!T$.
\item $T$ has a root: there exists exactly one finite path $\vec{s}\!=(s_{0})$ of length one in $T$. The state $s_{0}$, denoted by $root(T)$, is called the root of the tree $T$.
\end{enumerate}
 The set of \emph{children} of the node $\vec{s}$ in $T$ is the set $\{ \vec{t}\!\in\! T \ | \  \vec{t}=\vec{s}.s^{\prime} \ \wedge \ s^{\prime}\!\in\!S \}$.
We  consider the nodes $\vec{s}$ of $T$ as labeled by the $\mathit{last}$ function. 
\end{definition}

\begin{definition}[Uniquely and fully branching nodes of a tree]
A node $\vec{s}$ in a tree $T$, is said to be \emph{uniquely branching} in $T$
if either $last(\vec{s})\! \in\! S_{t}$ or $\vec{s}$ has a unique
child in $T$. Similarly, $\vec{s}$ is \emph{fully branching} in $T$
if, for every $s\! \in\! E(last(\vec{s}))$, it holds that $\vec{s}.s\!\in\!T$.
\end{definition}

\begin{definition}[Markov chain in $\arena$]
A \emph{Markov chain} in $\arena$ is a tree $M$  such that  for every every node $\vec{s} \!\in\! M$, the following conditions holds:
\begin{enumerate}[(1)]
\item If $last(\vec{s})\!\in\! S_{1}\cup S_{2}$  then $\vec{s}$ branches uniquely in $M$.
\item If $last(\vec{s})\!\in\! S_{N}$ then $\vec{s}$ branches fully in $M$.
\end{enumerate}
\end{definition}
Note that, since the set of $E$-successors of every state $s\!\in\!S_{N}$ is at most countable, every Markov chain in $\arena$ is a \emph{countably branching} tree with a countable set of nodes.

\begin{definition}[Probability measure $\mathbb{P}_{M}$]
\label{measure_definition}
Every Markov chain $M$ determines a probability assignment $\mathbb{P}_{M}(O_{\vec{s}})$ to every basic open set $O_{\vec{s}}\!\subseteq\!\mathcal{P}$, for $\vec{s}$ a  finite path $\vec{s}\!=\!(s_{0},s_{1}, ...,s_{n})$ with $n\!\in\!\mathbb{N}$,  defined as follows:

\begin{center}
 $\mathbb{P}_{M}(O_{\vec{s}}) \bydef	 \left\{     \begin{array}{l  l}  \displaystyle \prod \{ \pi (s_{i})(s_{i+1}) \ | \  0\!\leq\! i \!<\! n \  \wedge \ s_{i}\!\in\! S_{N}\}\ \  & $if $ \vec{s} \in M\\
 											 0 & $otherwise$
						 \end{array}      \right.$

\end{center} 
In other words, $\mathbb{P}_{M}$ assigns to the basic open set $O_{\vec{s}}\!\subseteq\!\mathcal{P}$, i.e, the set of all completed paths having $\vec{s}$ as prefix, value $0$ if $\vec{s}$ is not a path in $M$, and the product of all probabilities labeling the probabilistic steps in $\vec{s}$, otherwise. Note that if there are no probabilistic steps in $\vec{s}$, then $\mathbb{P}_{M}$ assigns to $O_{\vec{s}}$ probability $1$, which is the value of the empty product. The assignment $\mathbb{P}_{M}$ extends to a unique  probability measure on the Borel $\sigma$-algebra ($\mathcal{P},\Omega$) \cite{Tao_measuretheory}, which we also denote with $\mathbb{P}_{M}$.
\end{definition}

Given the previous definitions we can define the \emph{expected reward} of Player $1$ when a given Markov chain $M$ is the result (up to the behavior of Nature) of a play in the two-player stochastic game $\langle \arena,\Phi\rangle$.
\begin{definition}[Expected reward of $M$] Let $\langle \arena,\Phi\rangle$ be a $2\frac{1}{2}$-player game. We define the \emph{expected reward} of a Markov chain $M$ in $\arena$, denoted by $E(M)$, as follows: 
\begin{center}
$ E(M)=\displaystyle \int_{\mathcal{P} } \Phi \, \, d\  \mathbb{P}_{M}$.
\end{center}
This is a good definition because $\Phi$ is assumed to be Borel measurable, thus integrable.
\end{definition}

As usual in game theory, players' moves are determined by strategies. \\

\begin{definition}
An \emph{unbounded memory deterministic strategy} (or just a strategy) $\sigma_{1}$ for Player $1$ in $\arena$ is defined as a function $\sigma_{1}\!:\!\mathcal{P}_{1}^{<\omega}\!\rightarrow\! S\cup \{ \bullet  \}$ such that $\sigma_{1}(\vec{s})\!\in\! E(last(s))$ if $E(last(\vec{s}))\!\not =\! \emptyset$ and $\sigma_{1}(\vec{s})\!=\! \bullet$ otherwise. Similarly a strategy $\sigma_{2}$ for Player $2$ is defined as a function $\sigma_{2}\!:\!\mathcal{P}_{2}^{<\omega}\!\rightarrow\! S\cup \{ \bullet  \}$. We say that a strategy $\sigma_{1}$ for Player $1$ is \emph{memoryless}, if there exists a function $f \!:\! S_{1}\!\rightarrow\! S\cup \{ \bullet  \}$ such that for every $\vec{s}\!\in\! \mathcal{P}_{1}^{<\omega}$, the equality $\sigma_{1}(\vec{s})\!=\! f(last(\vec{s}))$ holds. Similarly, a strategy $\sigma_{2}$ for Player $2$ is memoryless if there exists a function $f \!:\! S_{2}\!\rightarrow\! S\cup \{ \bullet  \}$ such that for every $\vec{s}\!\in\! \mathcal{P}_{2}^{<\omega}$, the equality $\sigma_{2}(\vec{s})\!=\! f(last(\vec{s}))$ holds. In other words a strategy is memoryless if its decision on any history $\vec{s}$, only depends on the last state $last(\vec{s})$ of $\vec{s}$. A pair $\langle \sigma_{1},\sigma_{2}\rangle$ of strategies, one for each player, is called a  \emph{strategy profile} and determines the behaviors of both players.
\end{definition}

\begin{definition}[$M^{s_{0}}_{\sigma_{1},\sigma_{2}}$]
Given an initial state $s_{0}\!\in\! S$ and a strategy profile $\langle \sigma_{1},\sigma_{2}\rangle$, a unique Markov chain $M^{s_{0}}_{\sigma_{1},\sigma_{2}}$ is determined:
\begin{enumerate}[(1)]
\item the root of $M^{s_{0}}_{\sigma_{1},\sigma_{2}}$ is labeled with $s_{0}$,
\item for every $\vec{s}\!\in\! M^{s_{0}}_{\sigma_{1},\sigma_{2}}$, if $last(\vec{s})\!=\!s$ with $s\!\in\! S_{1}$ not a terminal state, then the unique child of  $\vec{s}$ in $M^{s_{0}}_{\sigma_{1},\sigma_{2}}$ is $\vec{s}.\big(\sigma_{1}(\vec{s})\big)$,
\item for every $\vec{s}\!\in\! M^{s_{0}}_{\sigma_{1},\sigma_{2}}$, if $last(\vec{s})\!=\!s$ with $s\!\in\! S_{2}$ not a terminal state, then the unique child of  $\vec{s}$ in $M^{s_{0}}_{\sigma_{1},\sigma_{2}}$ is $\vec{s}.\big(\sigma_{2}(\vec{s})\big)$.
\end{enumerate}
We denote with $\mathbb{P}^{s_{0}}_{\sigma_{1},\sigma_{2}}$ the probability measure $\mathbb{P}_{M^{s_{0}}_{\sigma_{1},\sigma_{2}}}$ over $\langle \mathcal{P},\Omega\rangle$ induced by the Markov chain $M^{s_{0}}_{\sigma_{1},\sigma_{2}}$.
\end{definition}

\begin{definition}\label{values_of_the_game}
Given a $2\frac{1}{2}$-player game $\mathcal{G}\!=\!\langle \arena,\Phi \rangle$  and an initial state $s\!\in\!S$, we define the \emph{lower value} and \emph{upper value} of the game $\mathcal{G}$ at $s$, denoted by $Val_{\downarrow}(\mathcal{G})(s)$ and $Val_{\uparrow}(\mathcal{G})(s)$ respectively, as follows:
\begin{center} 
$Val_{\downarrow}(\mathcal{G})(s) = \bigsqcup_{\sigma_{1}}  \bigsqcap_{\sigma_{2}} E( M^{s}_{\sigma_{1},\sigma_{2}}) \ \ \ \ \ Val_{\uparrow}(\mathcal{G})(s) = \bigsqcap_{\sigma_{2}}  \bigsqcup_{\sigma_{1}}E(M^{s}_{\sigma_{1},\sigma_{2}})$.
\end{center}
\end{definition}

$Val_{\downarrow}(\mathcal{G})(s)$ represents the (limit) expected reward  that Player $1$ can get, when the game begins at $s$, by choosing his strategy $\sigma_{1}$ first and then letting Player $2$ pick an appropriate counter strategy $\sigma_{2}$. Similarly $Val_{2}(\mathcal{G})(s)$ represents the (limit) expected reward that Player $1$ can get, when the game begins at $s$, by first letting Player $2$ choose a strategy $\sigma_{2}$ and then picking an appropriate counter strategy $\sigma_{1}$. Clearly $Val_{\downarrow}(\mathcal{G})(s)\leq Val_{\uparrow}(\mathcal{G})(s)$ for every $s\in S$.

\begin{definition}[$\epsilon$-optimal strategies]\label{eps_strategy}
Given a $2\frac{1}{2}$-player game  $\mathcal{G}\!=\!\langle \arena, \Phi \rangle$, a strategy $\sigma_{1}$ for Player $1$ is called \emph{$\epsilon$-optimal}, for some $\epsilon\!\in\![0,1]$, if the following inequality holds: 
\begin{center}$\bigsqcap_{\sigma_{2}} E( M^{s}_{\sigma_{1},\sigma_{2}}) >  Val_{\downarrow}(\mathcal{G})(s)- \epsilon$
\end{center}
for every game state $s$. Similarly a strategy $\sigma_{2}$ for Player $2$ is called \emph{$\epsilon$-optimal}, if the following inequality holds: 
\begin{center} 
$\bigsqcup_{\sigma_{1}} E(M^{s}_{\sigma_{1},\sigma_{2}}) <  Val_{\uparrow}(\mathcal{G})(s)+ \epsilon$
\end{center}
for every game state $s$. We refer to a strategy as \emph{optimal} if it is $0$-optimal. 
\end{definition}
Clearly, for every $\epsilon\! > \! 0$, there exist $\epsilon$-optimal strategies for Player $1$ and Player $2$. However, in general, there could be no optimal strategies, as stated in Proposition \ref{no-optimal} below.

\begin{definition} Given a $2\frac{1}{2}$-player game  $\mathcal{G}\!=\!\langle \arena, \Phi \rangle$, and an initial game state $s$, we say that the game  $\mathcal{G}$ is \emph{determined at $s$} if $Val_{\downarrow}(\mathcal{G})(s)= Val_{\uparrow}(\mathcal{G})(s)$. We say that the game $\mathcal{G}$ is \emph{determined} if it is determined at every  game state $s$.
\end{definition}

The following fundamental result is due to Donald\ A.\ Martin \cite{Martin98}.

\begin{theorem}[\cite{Martin98,MS98}]\label{blackwell_determinacy}
\label{determinacy} Every $2\frac{1}{2}$-player game $\mathcal{G}\!=\!\langle \arena,\Phi\rangle$ such that every state $s$ has at most countably many successor states, is determined.
\end{theorem}

In this paper we are interested in $2\frac{1}{2}$-player \emph{parity} games, which are  $2\frac{1}{2}$-player games $\langle \arena,\Phi\rangle$ whose payoff function $\Phi$ is induced by a \emph{parity structure}.
\begin{definition}
Given a $2\frac{1}{2}$-player arena $\arena\!=\! \langle (S,E),(S_{1},S_{2},S_{N}),\pi\rangle$, a \emph{parity structure} for $\arena$ is a pair  $\langle \Pr,\reward\rangle$ where $\Pr$ is called the \emph{priority assignment} and $\reward$ is a called the \emph{terminal reward assignment}. The priority assignment $\priority$ is a function $\priority\!:\! S\!\rightarrow\! \mathbb{N}$, such that the set $\Pr(S)\!=\!\{ n \ | \ \exists s\!\in\! S.\Pr(s)\!=\!n\}$ is finite. In other words $\Pr$ assigns to each state $s\!\in\!S$ a natural number, also referred to as a  \emph{priority}, taken from a finite pool of options $\{ n_{0},\dots, n_{k}\}\!=\! \Pr(S)$. We denote with $\max(\Pr)$, $\min(\Pr)$ and $|\Pr|$  the natural numbers $\max\{  n_{0},\dots, n_{k} \}$, $\min\{  n_{0},\dots, n_{k}\}$ and $|\{  n_{0},\dots, n_{k}\} |$ respectively. The terminal reward assignment $\reward$  is a function $\reward\! :\!S_{t}\!\rightarrow\! [0,1]$ assigning a value in the real interval $[0,1]$ to each terminal state $s\!\in\! S_{t}$. 
\end{definition}

\begin{definition}\label{WPR}
Let $\arena\!=\!\langle (S,E),(S_{1},S_{2},S_{N}),\pi\rangle$ be a $2\frac{1}{2}$-player arena and $\langle\Pr,\reward\rangle$  a parity structure for it. The payoff function $\Phi_{\langle\Pr,\reward\rangle}$ induced by $\langle\Pr,\reward\rangle$ is defined on every completed path $\vec{s}\!\in\!\mathcal{P}$ as follows:
\begin{enumerate}[(1)]
\item if $\vec{s}$ is a finite path, then $\Phi_{\langle\Pr,\reward\rangle}(\vec{s})= \reward\big(\textit{last}(\vec{s})\big)$,
\item if $\vec{s}$ is infinite, i.e., $\vec{s}\!=\! \{s_{i}\}_{i\in\mathbb{N}}$, then $\Phi_{\langle\Pr,\reward\rangle}(\vec{s})\!=\!1$ if the greatest priority assigned to infinitely many states $s_{i}$ in $\vec{s}$ is even, and $\Phi_{\langle\Pr,\reward\rangle}(\vec{s})=0$ otherwise. 
\end{enumerate}
\end{definition}
The payoff $\Phi_{\langle \priority, \reward \rangle}$ is Borel-measurable for every parity structure  $\langle \priority, \reward \rangle$ \cite{zielonka04}. 

\begin{definition}
A $2\frac{1}{2}$-player \emph{parity} game is  a  $2\frac{1}{2}$-player game $\mathcal{G}\!=\!\langle \arena,\Phi\rangle$ where $\Phi\!=\!\Phi_{\langle \priority, \reward \rangle}$ for some priority structure  $\langle \priority, \reward \rangle$ on $\arena$.
\end{definition}

The following fact about $2\frac{1}{2}$-player parity games is the main obstacle one encounter when trying to  extend the proof technique adopted in \cite{MM07}, for proving the equivalence  of the denotational and game semantics of pL$\mu$ under finite models, to arbitrary models.

\begin{proposition}[\cite{zielonka04}]\label{no-optimal}
There exists a $2\frac{1}{2}$-player parity game (with countably infinite state space $S$), such that no optimal strategy exists for either player. Moreover, no memoryless $\epsilon$-optimal strategy exists for either player.
\end{proposition}

Due to this technical issue, we shall prove the desired equivalence  by a different proof technique inspired by  the \emph{unfolding technique} of \cite{Gradel2007,Santocanale2002}. The following simple proposition will be used in Section \ref{main_proof}.

\begin{proposition}\label{fixed_point_proposition_1}
Let $\mathcal{G}\!=\!\langle \arena, \Phi_{\langle \priority, \reward \rangle}\rangle$ be a two player stochastic  parity game with arena $\arena\!=\! \langle (S,E),(S_{1},S_{2},S_{N}),\pi\rangle$. The functions $Val_{\downarrow}(\mathcal{G})$ and  $Val_{\uparrow}(\mathcal{G})$, of type $S\rightarrow[0,1]$, are fixed points of the  functional $\mathcal{F}\!:\![0,1]^{S}\!\rightarrow\![0,1]^{S}$ defined as follows:
\begin{center}
$\displaystyle \mathcal{F}(f)(s)= \left\{      \begin{array}{l  l}  	\reward(s)  & $if $E(s)\!=\!\emptyset$, i.e, if $s$ is a terminal state$ \\ 
							               	\displaystyle 	\bigsqcup_{t\in E(s)}f(t) & $if$ \  s\!\in\! S_{1}\\
									\displaystyle 	\bigsqcap_{t\in E(s)}f(t) & $if$ \  s\!\in\! S_{2}\\
									\displaystyle 	\sum_{t\in E(s)} \pi(s)(t) \cdot f(t) & $if$ \  s\!\in\! S_{N}\\
                      	      		     \end{array}      \right.$ 

\end{center}
\begin{proof}
The result easily follows from the fact, immediate to verify, that given any path $\vec{s}.\vec{t}\!\in\!\mathcal{P}_{\!\arena}$, the equality 
\begin{equation}\label{equation_prefix_independent}
 \Phi_{\langle \priority, \reward \rangle}(\vec{s}.\vec{t}) =  \Phi_{\langle \priority, \reward \rangle}(\vec{t})
 \end{equation} holds, i.e., the payoff assigned by $ \Phi_{\langle \priority, \reward \rangle}$ to a path in $\arena$ does not depend on any finite prefix of the path. We just prove that, for every $s\!\in\! S_{N}$, the equality
\begin{equation}\label{proposition1_eq1}
Val_{\downarrow}(\mathcal{G})(s)=\mathcal{F}\big( Val_{\downarrow}(\mathcal{G})\big)(s)
\end{equation}
holds. The other cases can be proved in a similar way.

Let $E(s)\!=\!\{t_{i}\}_{i\in I}$, for some (necessarily countable) index set $I$. By  definition of $Val_{\downarrow}(\mathcal{G})$, we need to prove that the equality 
\begin{equation}\label{proposition1_eq2}
 \displaystyle \bigsqcup_{\sigma_{1}} \bigsqcap_{\sigma_{2}} E(M^{s}_{\sigma_{1},\sigma_{2}})  =  \sum_{i\in I} \pi(s)(t_{i}) \cdot \big( \bigsqcup_{\sigma_{1}} \bigsqcap_{\sigma_{2}} E(M^{t_{i}}_{\sigma_{1},\sigma_{2}}) \big)
 \end{equation}
holds. This is done by proving the two inequalities ($\leq$) and ($\geq$) of Equation \ref{proposition1_eq2} separately. We just show how to prove the inequality ($\leq$) as the other one can be proved in a similar way. Assume, by contradiction, that the lefthand expression of  Equation \ref{proposition1_eq2} is strictly greater that the righthand expression. This means that there exists a strategy $\sigma_{1}$ for Player $1$ such that 
\begin{equation}\label{proposition1_eq3}
 \displaystyle \bigsqcap_{\sigma_{2}} E(M^{s}_{\sigma_{1},\sigma_{2}})  >  \sum_{i\in I} \pi(s)(t_{i})\cdot \big( \bigsqcup_{\sigma^{i}_{1}} \bigsqcap_{\sigma^{i}_{2}} E(M^{t_{i}}_{\sigma_{1},\sigma_{2}}) \big)
 \end{equation}
 holds. Since $s\!\in\! S_{N}$, i.e., the state $s$ is under the control of Nature, no action is made by either Player $1$ or Player $2$ at $s$, because the game progresses to some state $t\!\in\!E(s)$ accordingly with the random choice of Nature. Let us define, for every $i\!\in\! I$, the strategy $\tau^{i}_{1}$ for Player $1$ as follows: $\tau^{i}_{1}(\vec{s})=\sigma_{1}(s.\vec{s})$, for paths $\vec{s}$ starting at $t_{i}$. We do not need describe the behavior of $\tau^{i}_{1}$ at paths of different kind. Informally the strategy $\tau_{i}$, when the game starts at $t_{i}$, acts as the strategy $\sigma_{1}$ when the game starts at $s$ and Nature moves from $s$ to  $t_{i}$. The assumption of Equation \ref{proposition1_eq3} clearly implies that the following inequality
  \begin{equation}\label{proposition1_eq4}
 \displaystyle \bigsqcap_{\sigma_{2}} E(M^{s}_{\sigma_{1},\sigma_{2}})  > \sum_{i\in I} \pi(s)(t_{i})\cdot \big(  \bigsqcap_{\sigma^{i}_{2}} E(M^{t_{i}}_{\tau^{i}_{1},\sigma^{i}_{2}}) \big)
 \end{equation}
 holds. This in turn implies, since the set $I$ is countable, that there exist strategies $\{ \tau^{i}_{2}\}_{i\in I}$, such that the inequality 
\begin{equation}\label{proposition1_eq5}
 \displaystyle \bigsqcap_{\sigma_{2}} E(M^{s}_{\sigma_{1},\sigma_{2}})  > \sum_{i\in I} \pi(s)(t_{i})\cdot  E(M^{t_{i}}_{\tau^{i}_{1},\tau^{i}_{2}}) 
 \end{equation}
 holds. Let us define the strategy $\sigma_{2}$ for Player $2$ as follows: $\sigma_{2}(s.\vec{t})=\tau^{i}_{2}(\vec{t})$, for paths $\vec{t}$ starting at $t_{i}$, for $i\!\in\! I$. We do not need describe the behavior of $\sigma_{2}$ at paths of different kind, i.e., on paths not starting at $s$. Informally the strategy $\sigma_{2}$, when the game starts at $s$ and Nature randomly chooses to move to the state $t_{i}$, for $i\!\in\! I$,  play the rest of the game as the strategy $\tau^{i}_{2}$ would when the game starts at $t_{i}$. It then follows from Equation \ref{proposition1_eq5} that the following inequality
 \begin{equation}\label{proposition1_eq6}
 \displaystyle E(M^{s}_{\sigma_{1},\sigma_{2}})  > \sum_{i\in I} \pi(s)(t_{i})\cdot  E(M^{t_{i}}_{\tau^{i}_{1},\tau^{i}_{2}}) 
 \end{equation}
 holds. 
 
 We have just proved how the truth of Equation \ref{proposition1_eq6}  follows from the assumption of Equation \ref{proposition1_eq1}. We now derive the desired contradiction, by proving that Equation \ref{proposition1_eq6} does not hold because its two expression are equivalent.
 
 It follows immediately from the definition of the strategies $\{\tau^{i}_{1}\}_{i\in I}$ and $\sigma_{2}$, that the Markov chain $M^{s}_{\sigma_{1},\sigma_{2}}$ can be depicted as if Figure \ref{fig1_game_section}, 
 \begin{figure}
\centering
		\pstree[ treemode=U,levelsep=8ex ]{\Tr{$ s $}}{
        		\pstree[levelsep=5ex]{\Tr{$t_{i} $}  \tlput{$\lambda_{i}$} \trput{...} }{ 
		\pstree[linestyle=none,arrows=-,levelsep=3ex]{\Tfan[fansize=10ex]}{\TR{ $ M^{t_{i}}_{\tau^{i}_{1},\tau^{i}_{2}} $}}
		}
        		\pstree[levelsep=5ex]{\Tr{$ t_{j} $}  \trput{$\lambda_{j}$ ...}  }{ 
		\pstree[linestyle=none,arrows=-,levelsep=3ex]{\Tfan[fansize=10ex]}{\TR{ $ M^{t_{j}}_{\tau^{j}_{1},\tau^{j}_{2}} $}}
		}
	}
\caption{Markov chain $M^{s}_{\sigma_{1},\sigma_{2}}$}
 \label{fig1_game_section}
\end{figure}  
where the letters $i,j$ range over $I$, the value labeling the edge connecting $s$ with $t_{j}$ stands for $\pi(s)(t_{i})$ and highlights the fact that that edge is chosen by Nature with probability $\lambda_{i}$, and the subtree of $M^{s}_{\sigma_{1},\sigma_{2}}$ rooted at the state $t_{i}$ is precisely the Markov chain $M^{t_{i}}_{\tau^{i}_{1},\tau^{i}_{2}}$ induced by the strategies $\tau^{i}_{1}$ and $\tau^{i}_{2}$ at the starting state $t_{i}$.

It follows from Definition \ref{measure_definition} that the probability measure $\mathbb{P}_{M^{s}_{\sigma_{1},\sigma_{2}}}$ induced by  $M^{s}_{\sigma_{1},\sigma_{2}}$ assigns probability $0$ to the set of paths not starting at the state $s$. It then follows that the equality 
\begin{center} 
 $ E(M^{s}_{\sigma_{1},\sigma_{2}})\bydef\displaystyle \int_{\mathcal{P}_{\!\arena} } \Phi_{\langle \priority, \reward \rangle} \, \, d\  \mathbb{P}^{s}_{\sigma_{1},\sigma_{2}} = \displaystyle \sum_{i\in I} \big( \int_{O_{s.t_{i}} }\!\! \Phi_{\langle \priority, \reward \rangle} \, \, d\  \mathbb{P}^{s}_{\sigma_{1},\sigma_{2}}\big). $
\end{center}
holds, where $O_{s.t_{i}}$ denotes the open set of paths having $s$ and $t_{i}$ as first and second state respectively. Furthermore, again by Definition \ref{measure_definition}, the probability measure $\mathbb{P}_{M^{s}_{\sigma_{1},\sigma_{2}}}$ assigns probability $\pi(s)(t_{i})$ to the set $O_{s.t_{i}}$. From this observation, the previous considerations on the structure of $M^{s}_{\sigma_{1},\sigma_{2}}$ and its sub-Markov chains $M^{t_{i}}_{\tau^{i}_{1},\tau^{i}_{2}}$ and Equation \ref{equation_prefix_independent}, it follows immediately that the equality
\begin{center}
$\displaystyle \frac{1}{\pi(s)(t_{i})} \cdot  \big( \int_{O_{s.t_{i}} }\!\! \Phi_{\langle \priority, \reward \rangle} \, \, d\  \mathbb{P}^{s}_{\sigma_{1},\sigma_{2}}\big) =  \int_{\mathcal{P}} \Phi_{\langle \priority, \reward \rangle}  \, \, d\  \mathbb{P}^{t_{i}}_{\tau^{i}_{1},\tau^{i}_{2}} $ 
\end{center}
holds, where  $\mathbb{P}^{t_{i}}_{\tau^{i}_{1},\tau^{i}_{2}}$  is the probability measure over paths induced by the Markov chain $M^{t_{i}}_{\tau^{i}_{1},\tau^{i}_{2}}$. This concludes the proof.
\end{proof}
\end{proposition}

\section{Stochastic parity games for pL$\mu$}
\label{game_semantics}
In this section we define the \emph{game semantics} of the probabilistic modal $\mu$-calculus, in terms of $2\frac{1}{2}$-player parity games.

Given a PLTS $\langle P, \{ \freccia{a} \}_{a\in L}\rangle$, a pL$\mu$ formula $F$ and an interpretation $\rho$ of the variables, we denote with $\mathcal{G}^{F}_{\rho}$ the parity game $\langle \arena,\Phi_{\langle\priority,\reward\rangle}\rangle$ formally defined as follows. The state space of the arena $\arena\!=\!\langle (S,E), \{ S_{1},S_{2},S_{N} \},\pi\rangle$, is the set $S\!=\!  (P\times Sub(F)) \cup (\mathcal{D}(P)  \times Sub(F))$ of pairs of states $p\!\in\! P$ or probability distributions $d\!\in\!\mathcal{D}(P)$, and subformulas $G\!\in\! Sub(F)$. The transition relation $E$ is defined as $E (\langle d,G \rangle)\!=\!   \{ \langle p, G \rangle \ | \ p\!\in\! supp(d) \}$ for every probability distribution $d\!\in\! \mathcal{D}(P)$ and $E(\langle p, G\rangle)$, for $p\!\in\! P$,  is defined by case analysis on the outermost connective of $G$ as follows: 
\begin{enumerate}[(1)]
\item if $G= X$, with $X$ free in $F$, then $E(\langle p, G\rangle)\!=\!\emptyset$.
\item if $G=X$, with $X$ bound in $F$ by the subformula $\star X. H$, with $\star\! \in\! \{ \mu , \nu \}$, then $E(\langle p, G\rangle)\!=\! \{\langle p, H\rangle\}$.
\item if $G\!=\!\star X.H$, with $\star\! \in\! \{ \mu, \nu \}$, then $E(\langle p,G\rangle)\!=\! \{\langle p, H\rangle\}$.
\item if $G\!=\!\diam{a}H$ or $G\!=\!\quadrato{a}H$ then $E(\langle p, G\rangle) \!=\! \{ \langle d, H \rangle\ |\ p\freccia{a}d \}$. 
\item if $G\!=\! H\vee H^{\prime}$ or $G\!=\! H\wedge H^{\prime}$ then $E(\langle p, G\rangle) \!=\!\{ \langle p, H \rangle, \langle p,H^{\prime }\rangle \}$
\end{enumerate}
The partition $\{ S_{1},S_{2},S_{N} \}$ is defined as follows: every state $\langle p, G \rangle$ with $G$'s main connective in $\{ \diam{a}, \vee, \mu X\}$ or with $G\!=\!X$ where $X$ is a $\mu$-variable, is in $S_{1}$. Dually every state $\langle p, G\rangle$ with $G$'s main connective in $\{ \quadrato{a}, \wedge, \nu X\}$ or with $G\!=\!X$ where $X$ is a $\nu$-variable, is in $S_{2}$. Every state $\langle d, G \rangle$ is in $S_{N}$. Finally, the terminal states $\langle p, X\rangle$, with $X$ free in $F$, are in $S_{1}$ by convention. The probability transition function $\pi\!:\! S_{N}\!\rightarrow\! \mathcal{D}(S)$ is defined as $\pi (\langle d, G\rangle) (\langle p, G\rangle)\!=\!d(p)$. The priority assignment $\priority$ is defined as usual in $\mu$-calculus games (see, e.g., \cite{FGK2010}). The priority assigned to the states $\langle p, X \rangle$, with $X$ a $\mu$-variable,  is  a positive odd number; dually the priority assigned to the states $\langle p, X \rangle$, with $X$ a $\nu$-variable, is a positive even number. Moreover $\priority(\langle p, X\rangle) > \priority (\langle p^{\prime}, X^{\prime}\rangle )$ if $X$ subsumes $X^{\prime}$ in $F$. All other states get priority $0$. The terminal reward assignment $\reward$ is defined as $\reward(\langle p, X\rangle)=\rho(X)(p)$ for every terminal state $\langle p, X\rangle$ with $X$ free in $F$. All other terminal states in $\mathcal{G}^{F}_{\rho}$ are either of the form $\langle p, \diam{a}H\rangle$ or $\langle p, \quadrato{a}H\rangle$, with $\{ d \ | p\freccia{a}d\}\!=\!\emptyset$. The reward assignment $\reward$ is defined on these terminal states as follows: $\reward(\langle p, \diam{a}H\rangle)\!=\!0$ and $\reward(\langle p, \quadrato{a}H\rangle)\!=\!1$. This implements the policy that a player loses if they get stuck at these kind of states.

Observe that from the above definitions, in general, a pL$\mu$ game state $s\!\in\! {S_{1}\!\cup\! S_{2}}$ can have uncountably many $E$-successors. However the set $ supp(\pi(s))$ of $E$-successors of any state $s\!\in\! S_{N}$ is at most countable.

We are now ready to state our main theorem which asserts that every pL$\mu$-game is determined, and that the \emph{value} of the game at each state $\langle p, F\rangle$ coincides with the denotational interpretation $\sem{F}_{\rho}$ at $p$.

\begin{theorem} \label{teorema}
Given an arbitrary PLTS  $\langle P, \{ \freccia{a} \}_{a\in L}\rangle$, for every pL$\mu$ formula $F$, interpretation $\rho$ and process-state $p\!\in \!P$, the following equalities hold:
\begin{center} $\sem{F}_{\rho}(p) = Val_{\downarrow}(\mathcal{G}^{F}_{\rho})(\langle p, F\rangle) =  Val_{\uparrow}(\mathcal{G}^{F}_{\rho})(\langle p, F\rangle)$.
\end{center}
In particular  pL$\mu$ games are determined.
\end{theorem}

The proof of theorem \ref{teorema} is carried out in full detail in Section \ref{main_proof}.

\section{Proof of Theorem \ref{teorema}}\label{main_proof}

As anticipated in the introduction, the main difficulty in proving Theorem \ref{teorema} is that in general, as stated in Proposition \ref{no-optimal}, optimal strategies, or even memoryless $\epsilon$-optimal strategies may not exist in a given pL$\mu$-game. This compels us to use a different technique than the one adopted in, e.g., \cite{Stirling96,MM07}, which is based on the existence of  optimal memoryless strategies. Moreover, as observed earlier, since  pL$\mu$-games might have states with uncountably many successors, even the determinacy of  pL$\mu$-games does not follow directly from Theorem \ref{blackwell_determinacy}.

The proof technique we adopt is similar to the \emph{unfolding method} of \cite{FGK2010}. The unfolding method can be roughly described as a technique for proving \emph{properties} of (some sort of) two-player parity games by induction on the number of priorities used in the game. Usually, the first step is to prove that the property under consideration holds for all  parity games with just one priority. Then the the general result for games with $n+1$ priorities follows by some argument making use of the inductive hypothesis. In our setting we are interested in pL$\mu$-games of the form $\mathcal{G}^{F}_{\rho}$, and the property we want to prove is that the lower and upper values of these games coincide with the denotational value of $F$ under the interpretation $\rho$. We prove this by induction of the structure of $F$ rather than on the number of priorities used in the game $\mathcal{G}^{F}_{\rho}$. This allows a more transparent and arguably simpler proof.

More formally we shall prove, by induction on the structure of the formulas, that the following equations hold for every PLTS $\mathcal{L}\!=\!\langle P, \{ \freccia{a} \}_{a\in L}\rangle$, pL$\mu$ formula $F$ and $[0,1]$-interpretation $\rho$ of the variables:
\begin{equation}\label{main_equation_to_prove}
\forall G \in Sub(F), \ \ \sem{G}_{\rho}(p)=  Val_{\downarrow}( \mathcal{G}^{G}_{\rho})(\langle p, G\rangle) = Val_{\uparrow}( \mathcal{G}^{G}_{\rho})(\langle p, G\rangle).
\end{equation}

\textbf{Base case:} $\mathbf{G\!=\!X}$, for some variable $X\!\in\!Var$. \\
For every process state $p\!\in\!P$ and every interpretation $\rho$, the equality $\sem{X}_{\rho}(p)\!=\!\rho(X)(p)$ holds by Definition \ref{denotationa_sem_def}. In the game $\mathcal{G}^{X}_{\rho}$ the state $\langle p, X\rangle$ is \emph{terminal} (and therefore the game immediately terminates when starting at this state) and the terminal reward assignment $\reward$ is defined as $\reward(\langle p, X\rangle)\!=\!\rho(X)(p)$. The desired result (\ref{main_equation_to_prove}) then follows by application of Proposition \ref{fixed_point_proposition_1}.

\textbf{Inductive case} $\mathbf{G\!=\! G_{1} \vee G_{2}}$. \\
For every process state $p\!\in\!P$ and every interpretation $\rho$, we have by Definition \ref{denotationa_sem_def} that $\sem{G_{1}\vee G_{2}}_{\rho}(p)\!=\! \sem{G_{1}}_{\rho}(p) \sqcup \sem{G_{2}}_{\rho}(p)$ holds. Let us consider the state $\langle p, G_{1}\vee G_{2}\rangle$ of the game $\mathcal{G}^{G_{1}\vee G_{2}}_{\rho}$. This state is in $S_{1}$, i.e., under the control of Player $1$, which can choose to move either to $\langle p, G_{1}\rangle$ or $\langle p, G_{2}\rangle$. Observe that once the state $\langle p, G_{1}\vee G_{2}\rangle$ is left after the initial move, it is not reachable again in the game. Moreover, once  the state $\langle p, G_{i}\rangle$ is reached, $i\!\in\!\{1,2\}$, the rest of the game is identical to the game $\mathcal{G}^{G_{i}}_{\rho}$ (starting at $\langle p, G_{i}\rangle$). If follows from these observations that the equality
 \begin{equation}\label{vee_step_aux_1} 
Val_{\star}(\mathcal{G}^{G}_{\rho})(\langle p, G_{i}\rangle)=Val_{\star}(\mathcal{G}^{G_{i}}_{\rho})(\langle p, G_{i}\rangle)
\end{equation}
holds, for $i\!\in\!\{1,2\}$ and $\star\!\in\!\{\uparrow,\downarrow\}$.
By induction hypothesis we know that the equalities 
\begin{equation}\label{vee_step_aux_2}
\sem{G_{i}}_{\rho}(p)=Val_{\downarrow}( \mathcal{G}^{G_{i}}_{\rho})(\langle p, G_{i}\rangle) = Val_{\uparrow}( \mathcal{G}^{G_{i}}_{\rho})(\langle p, G_{i}\rangle)
\end{equation}
hold, for $i\!\in\!\{1,2\}$ and $\star\!\in\!\{\uparrow,\downarrow\}$. The desired result (\ref{main_equation_to_prove}) then follows immediately from equations \ref{vee_step_aux_1} and \ref{vee_step_aux_2} by application of Proposition \ref{fixed_point_proposition_1}.

\textbf{Inductive case} $\mathbf{G\!=\! G_{1} \wedge G_{2}}$. \\
Similar to the previous one.

\textbf{Inductive case} $\mathbf{G\!=\! \diam{a}G_{1}}$. \\
For every process state $p\!\in\!P$ and interpretation $\rho$, we have that the equality $\sem{\diam{a}G_{1}}_{\rho}(p)\!=\!  \displaystyle \bigsqcup_{p\freccia{a}d} \big( \sum_{q\in supp(d)} d(q)\! \cdot\! \sem{G_{1}}_{\rho}(q) \big)$ holds, by Definition \ref{denotationa_sem_def}.  Let us consider the state $\langle p, \diam{a}G_{1}\rangle$ of the game $\mathcal{G}^{\diam{a}G_{1}}_{\rho}$. This state is in $S_{1}$, i.e., under the control of Player $1$, which can move to a state in the (possibly empty) set $\{ \langle d, G_{1}\rangle \ | \ p\freccia{a}d \}$. As a first observation, note that if the set of $a$-successors of $p$, i.e., the set $\{ d \ | \ p\freccia{a}d\}$, is empty, then $\langle p, \diam{a}G_{1}\rangle$ is a terminal state and the terminal reward assignment $\reward$ is defined as $\reward(\langle p, \diam{a}G_{1}\rangle)\!=\!0$. The desired result (\ref{main_equation_to_prove}) then follows by Proposition \ref{fixed_point_proposition_1}. Let us then assume that  $\{ d \ | \ p\freccia{a}d\}\!=\!\{ d_{i} \}_{i\in I}$, for some non empty index-set $I$. Each state $\langle d_{i},G_{1}\rangle$ is under the control of Nature which moves to the state $\langle q, G_{1}\rangle$ with probability $d_{i}(q)$. More formally we have that $\langle d_{i},G_{1}\rangle\!\in\! S_{N}$ and $\pi \big( \langle d_{i},G_{1}\rangle \big)(\langle q, G_{1}\rangle)=d_{i}(q)$. Once the state  $\langle d_{i}, G_{1}\rangle$ is left, and the state $\langle q,G_{1}\rangle$ is reached, the rest of the game is identical to the game $\mathcal{G}^{G_{1}}_{\rho}$ (starting at $\langle q, G_{i}\rangle$), by considerations analogous to those discussed for the case $G\!=\!G_{1}\vee G_{2}$ above. If follows from this observation that the equality
 \begin{equation}\label{diam_step_aux_1} 
Val_{\star}(\mathcal{G}^{\diam{a}G_{1}}_{\rho})(\langle q, G_{1}\rangle)=Val_{\star}(\mathcal{G}^{G_{1}}_{\rho})(\langle q, G_{1}\rangle)
\end{equation}
holds for $\star\!\in\!\{\uparrow,\downarrow\}$.  By induction hypothesis we know that the equalities 
\begin{equation}\label{diam_step_aux_2}
\sem{G_{1}}_{\rho}(q)=Val_{\downarrow}( \mathcal{G}^{G_{1}}_{\rho})(\langle q, G_{1}\rangle) = Val_{\uparrow}( \mathcal{G}^{G_{1}}_{\rho})(\langle q, G_{1}\rangle)
\end{equation}
hold, where $\star\!\in\!\{\uparrow,\downarrow\}$, for every process state $q$. By applying twice the result of Proposition \ref{fixed_point_proposition_1} it then follows that
\begin{equation}
Val_{\star}(\mathcal{G}^{\diam{a}G_{1}}_{\rho})(\langle d_{i}, G_{1}\rangle) = \displaystyle \sum_{q\in supp(d)} d(q)\! \cdot\! \sem{G_{1}}_{\rho}(q) 
\end{equation}
for $\star\!\in\!\{\uparrow,\downarrow\}$ and every $d_{i}\!\in\! E(p)$, and
\begin{equation}
Val_{\star}(\mathcal{G}^{\diam{a}G_{1}}_{\rho})(\langle p,\diam{a} G_{1}\rangle)=  \displaystyle \bigsqcup_{p\freccia{a}d} \big( \sum_{q\in supp(d)} d(q)\! \cdot\! \sem{G_{1}}_{\rho}(q) \big)
\end{equation}
hold as desired. 

\textbf{Inductive case} $\mathbf{G\!=\! \quadrato{a}G_{1}}$. \\
Similar to the previous one.

\textbf{Inductive case} $\mathbf{G\!=\! \mu X.G_{1}}$. \\
For every process state $p$ and every interpretation $\rho$ we have, by Definition \ref{denotationa_sem_def}, that the following equality holds:
\begin{center}
$\sem{\mu X.G_{1}}_{\rho}(p)\bydef \lfp \Big( \lambda f\!\in\![0,1]^{P}. \big( \sem{G_{1}}_{\rho[f/X]}\big)  \Big)(p)$.
\end{center}
By the Knaster-Tarski theorem, the previous equation can be rewritten as:
\begin{equation}\label{fix_step_aux_1}
\sem{\mu X.G_{1}}_{\rho}(p)=\bigsqcup_{\alpha} \sem{G_{1}}^{\alpha}_{\rho},
\end{equation}
where $\alpha$ ranges over the ordinals, and $\sem{G_{1}}^{\alpha}_{\rho}$ is defined as $\bigsqcup_{\beta<\alpha}\sem{G_{1}}_{\rho[\sem{G_{1}}^{\beta}_{\rho}/X]}$. Let us denote with $\gamma$ the least ordinal such that $\sem{G_{1}}^{\gamma}_{\rho}=\sem{\mu X.G_{1}}_{\rho}$, and with $\rho_{\gamma}\!\in\![0,1]^{P}$ the interpretation $\rho[\sem{G_{1}}^{\gamma}_{\rho}/X]$. Thus,  the following equation holds: 
\begin{equation}
\sem{G_{1}}_{\rho_{\gamma}}=\sem{\mu X.G_{1}}_{\rho}.
\end{equation}

Let us now turn our attention to the $2\frac{1}{2}$-player parity game $\mathcal{G}^{\mu X.G_{1}}_{\rho}$. Our goal is to prove Equation \ref{main_equation_to_prove}, i.e., that the following equalities
\begin{equation}\label{fix_step_aux_2}
\sem{\mu X.G_{1}}_{\rho}(p) = Val_{\downarrow}\big(\mathcal{G}^{\mu X.G_{1}}_{\rho}\big)(\langle p, \mu X.G_{1}\rangle)= Val_{\uparrow}\big(\mathcal{G}^{\mu X.G_{1}}_{\rho}\big)(\langle p, \mu X.G_{1}\rangle)
\end{equation}
hold, for every $p\!\in\!P $. As a first observation, note that the state $\langle p, \mu X.G_{1}\rangle$ is not reachable by any other game state, and that it has the state $\langle p, G_{1}\rangle$ as its only successor state. It then follows by application of Proposition \ref{fixed_point_proposition_1} that, in order to prove the desired result (\ref{fix_step_aux_2}), we just have to show that the equalities 
\begin{equation}\label{fix_step_aux_3}
\sem{G_{1}}_{\rho_{\gamma}}(p) = Val_{\downarrow}\big(\mathcal{G}^{\mu X.G_{1}}_{\rho}\big)(\langle p,G_{1}\rangle)= Val_{\uparrow}\big(\mathcal{G}^{\mu X.G_{1}}_{\rho}\big)(\langle p, G_{1}\rangle)
\end{equation}
hold. In order to improve readability, we shall denote with $\gsem{\mu X.G_{1}}^{\star}_{\rho}\!:\!P\!\rightarrow\![0,1]$ the function defined as $\lambda p\!\in\!P.\Big( Val_{\star}\big(\mathcal{G}^{\mu X.G_{1}}_{\rho}\big)(\langle p,G_{1}\rangle) \Big)$, for $\star\!\in\!\{\downarrow,\uparrow\}$. Thus, Equation \ref{fix_step_aux_3} can be rewritten as follows:
\begin{equation}\label{fix_step_aux_3_prime}
\sem{G_{1}}_{\rho_{\gamma}}(p) = \gsem{\mu X.G_{1}}^{\downarrow}_{\rho}(p)=\gsem{\mu X.G_{1}}^{\uparrow}_{\rho}(p).
\end{equation}
Note that the analogous functions $\gsem{G_{1}}^{\star}_{\rho[f/X]}\!:\!P\!\rightarrow\![0,1]$ specified, for $\star\!\in\!\{\downarrow,\uparrow\}$, as \\ $\lambda p\!\in\!P.\Big( Val_{\star}\big(\mathcal{G}^{G_{1}}_{\rho[f/X]}\big)(\langle p,G_{1}\rangle) \Big)$, satisfy the following equation:
\begin{equation}\label{fix_step_aux_3_second}
\sem{G_{1}}_{\rho[f/X]}= \gsem{G_{1}}^{\downarrow}_{\rho[f/X]} = \gsem{G_{1}}^{\uparrow}_{\rho[f/X]} 
\end{equation}
for all $f\!\in\![0,1]^{P}$, by induction hypothesis on $G_{1}$.

We prove Equation \ref{fix_step_aux_3_prime} by exploiting the similarities between the game $\mathcal{G}^{\mu X.G_{1}}_{\rho}$ and the game $ \mathcal{G}^{G_{1}}_{\rho[f/X]}$, for every $f\!\in\![0,1]^{P}$. The two games are indeed almost identical and differ only in the following two points:
\begin{enumerate}[(1)]
\item the set of game states of $\mathcal{G}^{\mu X.G_{1}}_{\rho}$ consists of the game states of $ \mathcal{G}^{G_{1}}_{\rho[f/X]}$ plus the set $\{ \langle p, \mu X.G_{1}\rangle \ | \ p\!\in\! P\}$,
\item the states of the form $\langle p, X\rangle$, for $p\!\in\!P$, are terminal in the game $ \mathcal{G}^{G_{1}}_{\rho[f/X]}$ and, instead,  have the state $\langle p, G_{1}\rangle$ as unique successor in $\mathcal{G}^{\mu X.G_{1}}_{\rho}$.
\end{enumerate}
The first point does not contribute to any significant difference between the two games, because, as already observed earlier, the states of the form $\langle p, \mu X.G_{1}\rangle$, for $p\!\in\!P$, have a unique child and once left are not reachable again in the game, and hence can be ignored completely. Thus, in what follows, we will assume that the two games $\mathcal{G}^{\mu X.G_{1}}_{\rho}$ and $ \mathcal{G}^{G_{1}}_{\rho[f/X]}$ have the same set of states.
The second point is, on the other hand, an important one. In the game $ \mathcal{G}^{G_{1}}_{\rho[f/X]}$, when a state of the form $\langle p, X\rangle$ is reached, the play ends with  reward $f(p)$ for Player $1$. In the game $\mathcal{G}^{\mu X.G_{1}}_{\rho}$, instead, the game progresses to the state $\langle p, G_{1}\rangle$ and, from there, continues.

Given these observations it is clear that any finite path in  $ \mathcal{G}^{G_{1}}_{\rho[f/X]}$ is also a finite path in $\mathcal{G}^{\mu X.G_{1}}_{\rho}$. Moreover we define the functions $count$ and  $tail$ from finite paths in  $\mathcal{G}^{\mu X.G_{1}}_{\rho}$ to natural numbers and finite paths in  $ \mathcal{G}^{G_{1}}_{\rho[f/X]}$ respectively, as follows:
\begin{center}
$count(\vec{s})= \left| \{ \langle q, X\rangle.\langle q,G_{1}\rangle   \in \vec{s} \ | \ q\!\in\! P\}\right|$
\end{center}
and
\begin{center}
$tail(\vec{s}) = \left\{     \begin{array}{l  l}  	
\vec{s} & $  if $ count(\vec{s})\!=\!0   \\
\vec{t} & $ if $\vec{s}\!=\! \vec{s^{\prime}}.\langle q,X\rangle.\vec{t} $ and $count(\vec{t})=0 $, for $q\!\in\!P

                      	      		     \end{array}      \right.$
			     \end{center}		
In other words $count(\vec{s})$ gives us the number of occurrences of pairs of (adjacent) states of the form  $\langle q, X\rangle$ and $\langle q, G_{1}\rangle$ in $\vec{s}$, for $q\!\in\!P$, and the finite path $tail(\vec{s})$ is obtained by removing from $\vec{s}$ the initial prefix up to the last occurrence of a state of the form $\langle q, X\rangle$ (immediately followed by the state $\langle q, G_{1}\rangle$) in $\vec{s}$. Note that $tail(\vec{s})$ is indeed a finite path in  $ \mathcal{G}^{G_{1}}_{\rho[f/X]}$. The function $count$ extends to an operation from completed (i.e., either terminated or infinite) paths to $\mathbb{N}\cup\{\infty\}$ as expected. Similarly, we extend the function $tail$ to an operation from infinite paths $\vec{s}$ in  $\mathcal{G}^{\mu X.G_{1}}_{\rho}$ (such that $count(\vec{s})\!<\!\infty$) to infinite paths in  $ \mathcal{G}^{G_{1}}_{\rho[f/X]}$, in the obvious way.

As a further remark about the similarities between the two games $\mathcal{G}^{\mu X.G_{1}}_{\rho}$ and $ \mathcal{G}^{G_{1}}_{\rho[f/X]}$, observe that the priorities assigned to the states of the two games coincide (or at least the can be made to coincide) except that the states of the form $\langle p, X\rangle$ are assigned priority $0$ in $ \mathcal{G}^{G_{1}}_{\rho[f/X]}$ and  maximal priority in $\mathcal{G}^{\mu X.G_{1}}_{\rho}$. Similarly, the terminal reward assignments of the two games coincide on all terminal states except that, on those of the form $\langle p, X\rangle$, the reward assignment of $\mathcal{G}^{\mu X.G_{1}}_{\rho}$  is not defined since because, as observed before, $\langle p, X\rangle$ is not a terminal state in $\mathcal{G}^{\mu X.G_{1}}_{\rho}$. It is then simple to verify that the following property holds for every  completed  path $\vec{s}$ in $\mathcal{G}^{\mu X.G_{1}}_{\rho}$:
\begin{equation}\label{fix_step_aux_4}
\Phi^{\mu X.G_{1}}_{\rho}(\vec{s})= \left\{ \begin{array}{l  l}  	
0 & $if $count(\vec{s})=\infty  \\
  \Phi^{G_{1}}_{\rho[f/X]}\big(tail(\vec{s})\big) & $otherwise$
 \end{array}      \right.
\end{equation}
where $\Phi^{\mu X.G_{1}}_{\rho}$ and $\Phi^{G_{1}}_{\rho[f/X]}$ denote the payoff functions of the two games $\mathcal{G}^{\mu X.G_{1}}_{\rho}$
 and $\mathcal{G}^{G_{1}}_{\rho[f/X]}$ 
 respectively.
The first clause of Equation \ref{fix_step_aux_4} holds because the priority assigned to  states of the form $\langle p, X\rangle$ is odd and maximal in $\mathcal{G}^{\mu X.G_{1}}_{\rho}$. The second clause follows immediately by previous observations.

One last observation, which follows immediately from previous considerations about the similarities between the two games $\mathcal{G}^{\mu X.G_{1}}_{\rho}$ and $\mathcal{G}^{G_{1}}_{\rho[f/X]}$, is the following: 
\begin{equation}\label{fix_step_aux_6}
\gsem{\mu X.G_{1}}^{\star}_{\rho} =  \gsem{G_{1}}^{\star}_{\rho[\gsem{\mu X.G_{1}}^{\star}_{\rho}/X]} \stackrel{\textnormal{Eq. \ref{fix_step_aux_3_second}}}{=} \sem{G_{1}}_{\rho[\gsem{\mu X.G_{1}}^{\star}_{\rho}/X]}
\end{equation}
for $\star\!\in\!\{\downarrow,\uparrow\}$. By application of Equation \ref{fix_step_aux_3_second}, this implies that both $\gsem{\mu X.G_{1}}^{\uparrow}_{\rho}$ and $\gsem{\mu X.G_{1}}^{\downarrow}_{\rho}$ are fixed points of $\lambda f\!\in\![0,1]^{P}.( \sem{G_{1}}_{\rho[f/X]})$. Note that, for all $p\!\in\!P$, the inequality $\gsem{\mu X.G_{1}}^{\downarrow}_{\rho}(p) \leq \gsem{\mu X.G_{1}}^{\uparrow}_{\rho}(p)$ holds. Moreover the inequality $\sem{\mu X.G_{1}}_{\rho}(p) \leq \gsem{\mu X.G_{1}}^{\downarrow}_{\rho}(p)$ hods, for all $p\!\in\!P$, because $\sem{\mu X.G_{1}}_{\rho}$ (or, equivalently, $\sem{G_{1}}_{\rho_{\gamma}}$) is the least fixed point of $\lambda f\!\in\![0,1]^{P}.( \sem{G_{1}}_{\rho[f/X]})$.

We shall prove the desired result (Equation \ref{fix_step_aux_3_prime}) by showing that, for all $p\!\in\!P$, the inequality
\begin{equation}
\gsem{\mu X.G_{1}}^{\uparrow}_{\rho}(p)\bydef Val^{\uparrow}\big( \mathcal{G}^{\mu X.G_{1}}_{\rho}\big)(\langle p, G_{1}\rangle) \leq   \sem{G_{1}}_{\rho_{\gamma}}(p)
\end{equation}
holds. We do this by constructing, for every $\epsilon\!>\!0$ and for every $k\!\in\!\mathbb{N}$, a strategy $\sigma^{[k]}_{2}$ for Player $2$ in the game $\mathcal{G}^{\mu X.G_{1}}_{\rho}$, satisfying the following inequality:
 \begin{equation}\label{strategy_eq_1}
  \bigsqcup_{\sigma_{1}}E(M^{\langle p, G_{1}\rangle}_{\sigma_{1},\sigma^{[k]}_{2}}) <   \sem{G_{1}}_{\rho_{\gamma}}(p) + \frac{\epsilon}{2^{k}} .
  \end{equation}
 Let us fix an arbitrary $\epsilon\!>\!0$. In what follows, we adopt the convention of using $\sigma$ and $\tau$ to range over strategies in the games $\mathcal{G}^{\mu X.G_{1}}_{\rho}$ and $\mathcal{G}^{G_{1}}_{\rho_{\gamma}}$ respectively.  The strategy $\sigma^{[k]}_{2}$, for $k\!\in\!\mathbb{N}$, is built using the collection of $\delta$-optimal strategies $\tau^{\delta}_{2}$, with $\delta \! >\!0$, for Player $2$ in the game $\mathcal{G}^{G_{1}}_{\rho_{\gamma}}$, i.e., strategies $\tau_{2}^{\delta}$ such that the inequality
\begin{equation} \label{delta_optimal_strat_1}
 \bigsqcup_{\tau_{1}}M^{\langle q, G_{1}\rangle}_{\tau_{1},\tau^{\delta}_{2}} \!< \! Val_{\uparrow}(\mathcal{G}^{G_{1}}_{\rho_{\gamma}})(\langle q,G_{1}\rangle)+\delta
 \end{equation}
 holds, for every $q\!\in\! P$. The strategy $\sigma_{2}^{[k]}$ is defined as follows:
\begin{center}
$\sigma^{[k]}_{2}(\vec{s}) = \left\{     \begin{array}{l  l}  	
\tau^{\frac{\epsilon}{2^{k+1}}}_{2}(\vec{s}) & $ if $  count(\vec{s})\!=\!0 \\
\sigma^{[k+i]}_{2}(\vec{t}) & $ if $  count(\vec{s})\!=\!i\!>\!0 $ and $ \vec{t}\!=\!tail(\vec{s})
 \end{array}      \right.$
\end{center}		
where the function $count$ and $tail$ have been defined earlier.
The strategy $\sigma_{2}^{[k]}$ can be informally described as follows: at the beginning of the game, $\sigma^{[k]}_{2}$  initially behaves as the strategy $\tau^{\frac{\epsilon}{2^{k+1}}}_{2}$. If a state of the form $\langle q, X\rangle$,  for $q\!\in\! P$, is ever reached, then Player $2$ forgets the previous game-history and improves their strategy behaving, from the subsequent state $\langle  p, G_{1}\rangle$, as the strategy $\sigma^{[k+1]}_{2}$. Further changes of strategy, from $\sigma^{[i]}_{2}$ to $\sigma^{[i+1]}_{2}$, for $i\!\in\!\mathbb{N}$, are repeated every time a state of the form $\langle q^{\prime}, X\rangle$ is reached, for $q^{\prime}\!\in\!P$. This means that on a history of the form $\vec{s}\!=\!\vec{s^{\prime}}.\langle q, X\rangle.\vec{t}$, where  $\langle q, X\rangle$ is the last occurrence of a state of the form $\langle q^{\prime}, X\rangle$ in $\vec{s}$, the choice of $\sigma_{2}^{[k]}$ at $\vec{s}$  coincides with that of $\sigma_{2}^{[k+i]}$ (or equivalently with $\tau^{\frac{\epsilon}{2^{k+i+1}}}_{2}$) at $\vec{t}$, where $i\!=\! count(\vec{s})$. 

In other words Player $2$, using the strategy $\sigma_{2}^{[k]}$, plays in $\mathcal{G}^{\mu X.G_{1}}_{\rho}$ as if they were playing in the game $\mathcal{G}^{G_{1}}_{\rho_{\gamma}}$, and every time a state of the form $\langle q, X\rangle$ is reached, they re-start again (from the unique successor $\langle q,G_{1}\rangle$ of $\langle q, X\rangle$) as if they ware in  $\mathcal{G}^{G_{1}}_{\rho_{\gamma}}$, but with an improved strategy. 

We now prove  that, for every $k\!\in\!\mathbb{N}$, the strategy $\sigma^{[k]}_{2}$ satisfies the desired Inequality \ref{strategy_eq_1}. Let us fix an arbitrary strategy $\sigma_{1}$ for Player $1$ in the game $\mathcal{G}^{\mu X.G_{1}}_{\rho}$. We just need to show that the inequality  
\begin{equation}\label{strategy_eq_2}
E(M^{\langle p, G_{1}\rangle}_{\sigma_{1},\sigma^{[k]}_{2}}) \!<\! \sem{G_{1}}_{\rho_{\gamma}}(p) + \frac{\epsilon}{2^{k}}
\end{equation}
 holds. Let us denote with $\mathcal{X}^{n}$, for $n\!\in\!\mathbb{N}$, the sets of completed paths $\vec{s}$ in $\mathcal{G}^{\mu X.G_{1}}_{\rho}$ such that $count(\vec{s})\!=\!n$. Let $\mathcal{X}^{\prec n}$  be the set $\bigcup_{i\prec n} \mathcal{X}^{i}$, for $\prec\in\{<,\leq\}$. Similarly, let us denote with $\mathcal{X}^{\infty}$ the set of completed paths $\vec{s}$ in $\mathcal{G}^{\mu X.G_{1}}_{\rho}$ such that $count(\vec{s})\!=\!\infty$.  The following equalities hold:
\begin{center}
\begin{tabular}{l l l }
$E(M^{\langle p, G_{1}\rangle}_{\sigma_{1},\sigma^{[k]}_{2}})$ & $\bydef$ &  $\displaystyle \int_{\mathcal{P}} \Phi^{\mu X.G_{1}}_{\rho}\ d \mathbb{P}^{\langle p, G_{1}\rangle}_{\sigma_{1},\sigma^{[k]}_{2}}$\\
$$ & $=_{A}$ & $\displaystyle \sum_{n\in\mathbb{N}} \int_{\mathcal{X}^{n}} \Phi^{\mu X.G_{1}}_{\rho}\ d \mathbb{P}^{\langle p, G_{1}\rangle}_{\sigma_{1},\sigma^{[k]}_{2}} + \int_{\mathcal{X}^{\infty}} \Phi^{\mu X.G_{1}}_{\rho}\ d \mathbb{P}^{\langle p, G_{1}\rangle}_{\sigma_{1},\sigma^{[k]}_{2}}$  \\
$$ & $=_{B}$ & $\displaystyle \sum_{n\in\mathbb{N}} \int_{\mathcal{X}^{n}} \Phi^{\mu X.G_{1}}_{\rho}\ d \mathbb{P}^{\langle p, G_{1}\rangle}_{\sigma_{1},\sigma^{[k]}_{2}}$ \\
$$ & $=_{C}$ & $\displaystyle\bigsqcup_{n\in\mathbb{N}}\displaystyle \int_{\mathcal{X}^{\leq n}} \Phi^{\mu X.G_{1}}_{\rho}\ d \mathbb{P}^{\langle p, G_{1}\rangle}_{\sigma_{1},\sigma^{[k]}_{2}}$ \\
\end{tabular}
\end{center}
The validity of equations $A$ and $C$ comes from countable additivity and $\omega$-continuity of the probability measure $ \mathbb{P}^{\langle p, G_{1}\rangle}_{\sigma_{1},\sigma^{[k]}_{2}}$ respectively, and the validity of equation $B$ follows from the fact that $\Phi^{\mu X.G_{1}}_{\rho}(\vec{s})\!=\!0$,  for every $\vec{s}\in \mathcal{X}^{\infty}$ (see Equation \ref{fix_step_aux_4}).

We now prove, by induction on the natural numbers, that for every $n\!\in\!\mathbb{N}$  the  inequality 
\begin{equation}\label{strategy_eq_3}
   \displaystyle \int_{\mathcal{X}^{\leq n}} \Phi^{\mu X.G_{1}}_{\rho}\ d \mathbb{P}^{\langle p, G_{1}\rangle}_{\sigma_{1},\sigma^{[k]}_{2}}  <  \sem{G_{1}}_{\rho_{\gamma}}(p) + \displaystyle \sum_{i\leq n} \frac{\epsilon}{2^{k+i+1}}
\end{equation}
holds. This clearly implies the desired Inequality \ref{strategy_eq_2} because the indentiy $\displaystyle\bigsqcup_{n\in\mathbb{N}}\displaystyle \sum_{i\leq n} \frac{\epsilon}{2^{k+i+1}}= \frac{\epsilon}{2^{k}}$ holds. Suppose, by inductive hypothesis, that the inequality (\ref{strategy_eq_3}) holds for all $m\!<\!n$, for some $n\!\in\!\mathbb{N}$. The Markov chain $M^{\langle p, G_{1}\rangle}_{\sigma_{1},\sigma^{[k]}_{2}}$ can be depicted as in figure \ref{fig4} where the triangle (denoted by $\mathcal{X}^{0}$) represents the set of paths in $M^{\langle p, G_{1}\rangle}_{\sigma_{1},\sigma^{[k]}_{2}}$ never reaching a state of the form $\langle q,X\rangle$, for $q\!\in\!P$, and the finite paths (denoted by $\vec{t}_{i}$) connecting the root $\langle p, G_{1}\rangle$ with the node $\langle q_{i}, X\rangle$,  for $i\!\in\! I\subseteq\! \mathbb{N}$, are the prefixes (up to the first occurrence of a state of the form $\langle q_{i},X\rangle$) of all paths $\vec{s}$ in $M^{\langle p, G_{1}\rangle}_{\sigma_{1},\sigma^{[k]}_{2}}$ of the form $\vec{s}\!=\!\vec{t_{i}}.\langle q_{i},G_{1}\rangle.\vec{s^{\prime}}$. The sub-Markov chains rooted at $\vec{t}_{i}$ (having $\langle q_{i},G_{1}\rangle$ as initial state) are denoted by $M_{i}$, for $i\!\in\!I$. 
\begin{figure}
\centering

\pstree[ treemode=U,levelsep=8ex ]{\Tr{$\langle p, G_{1}\rangle$}}{
        \pstree[linestyle=none,arrows=-,levelsep=2ex]{\Tfan[fansize=10ex]}{\TR{ $ \mathcal{X}^{0} $}}
  
\pstree[levelsep=5ex]{\Tr{$\langle q_{0},X\rangle$} \tlput{$\vec{t}_{0}$}\trput{$...$}}{
	\pstree[levelsep=5ex]{\Tr{$\langle q_{0},G_{1}\rangle$}}{
		\pstree[linestyle=none,arrows=-,levelsep=2ex]{\Tfan[fansize=10ex]}{\TR{ $M_{0} $}}
	
	}
}

\pstree[levelsep=5ex]{\Tr{$\ \ \ \ \langle q_{i},X\rangle\ \dots $} \trput{$\vec{t}_{i}\ ...$}}{
	\pstree[levelsep=5ex]{\Tr{$\langle q_{i},G_{1} \rangle$}}{
		\pstree[linestyle=none,arrows=-,levelsep=2ex]{\Tfan[fansize=10ex]}{\TR{ $M_{i} $}}
	}
}
}	
\caption{Markov chain $M^{\langle p, G_{1}\rangle}_{\sigma_{1},\sigma^{[k]}_{2}}$ in  $\mathcal{G}^{\mu X.G_{1}}_{\rho}$}
 \label{fig4}
\end{figure}

Note that every path $\vec{s}\!\in\!\mathcal{X}^{\leq n}$ is either a path in $\mathcal{X}^{0}$, i.e., does not have any occurrences of states of the form $\langle q, X\rangle$, or is in $\bigcup_{0<j\leq n}\mathcal{X}^{j}$. Moreover observe that any path $\vec{s}\!\in\!\bigcup_{0<j\leq n}\mathcal{X}^{j}$ in $M^{\langle p, G_{1}\rangle}_{\sigma^{\epsilon}_{1},\sigma^{[k]}_{2}}$, i.e., any path in $M^{\langle p, G_{1}\rangle}_{\sigma^{\epsilon}_{1},\sigma^{[k]}_{2}}$ that reaches at least once (and at most $n$ times) a state of the form $\langle q, X\rangle$, can be uniquely written as the concatenation $\vec{s}\!=\!\vec{t_{i}}.\vec{s^{\prime}}$ of some finite path $\vec{t}_{i}$ (ending in the state $\langle q_{i}, X\rangle$, which is the first occurrence of a state of this shape in $\vec{s}$) and some completed path $\vec{s^{\prime}}\!\in\! \mathcal{X}^{< n}$, which is necessarily starting at the state $\langle q_{i}, G_{1}\rangle$.   Let us denote with $\vec{t_{i}}.\mathcal{X}^{<n}$, for $i\!\in\!I$, the set of paths $\vec{s}\!\in\! \bigcup_{0<j\leq n}\mathcal{X}^{j}$ of the form $\vec{t_{i}}.\vec{s^{\prime}}$, with $\vec{s^{\prime}}\!\in\!  \mathcal{X}^{< n} $. Given the previous observations, and since the set $I$ is countable, the following equality holds:
\begin{equation}\label{n1_a}
\displaystyle   \int_{\mathcal{X}^{\leq n} }  \!\!\Phi^{\mu X.G_{1}}_{\rho} \, \, d \mathbb{P}^{\langle p, G_{1}\rangle}_{\sigma_{1},\sigma^{[k]}_{2}}= \displaystyle   \int_{\mathcal{X}^{0} } \!\! \Phi^{\mu X.G_{1}}_{\rho} \, \, d \mathbb{P}^{\langle p, G_{1}\rangle}_{\sigma_{1},\sigma^{[k]}_{2}} + \displaystyle \sum_{i\in I}   \int_{\vec{t}_{i}.\mathcal{X}^{< n} } \!\! \Phi^{\mu X.G_{1}}_{\rho}\, \, d \mathbb{P}^{\langle p, G_{1}\rangle}_{\sigma_{1},\sigma^{[k]}_{2}} 
\end{equation}
Moreover, denoting by $\pi(\vec{t}_{i})$, for $i\!\in\!I$, the multiplication of all probabilities appearing in the probabilistic steps of the path $\vec{t}_{i}$, it is simple to check that the following equality holds:
\begin{equation}\label{n2_a}
 \displaystyle \int_{\vec{t}_{i}.\mathcal{X}^{< n} } \!\! \Phi^{\mu X.G_{1}}_{\rho} \, \, d \mathbb{P}^{\langle p, G_{1}\rangle}_{\sigma_{1},\sigma^{[k]}_{2}}=  \pi(\vec{t}_{i}) \cdot\displaystyle \int_{\mathcal{X}^{< n} } \!\! \Phi^{\mu X.G_{1}}_{\rho} \, \, d \mathbb{P}_{M_{i}}
\end{equation}
where $\mathbb{P}_{M_{i}}$ denotes the probability measure over completed paths induced by the sub-Markov chain $M_{i}$. 

It follows from the definition of the strategy $\sigma^{[k]}_{2}$, that the sub-Markov chain $M_{i}$, for $i\!\in\!I$, is generated by the strategy profile $\langle \sigma^{i}_{1}, \sigma^{[k+1]}_{2}\rangle$, where $\sigma^{i}_{1}(\vec{s})\bydef\sigma_{1}(\vec{t}_{i}.\vec{s})$, for all completed paths $\vec{s}$ having $\langle q_{i},G_{1}\rangle$ as first state. Thus, $M_{i}=M^{q_{i}}_{\sigma^{i}_{1},\sigma^{[k+1]}_{2}}$. It then follows by inductive hypothesis on $n$ (Inequality \ref{strategy_eq_3}), that the inequality
\begin{equation}
 \displaystyle \int_{\mathcal{X}^{< n} } \!\!\Phi^{\mu X.G_{1}}_{\rho} \, \, d\mathbb{P}_{M_{i}} <  \sem{G_{1}}_{\rho_{\gamma}}(q_{i}) + \displaystyle \sum_{j<n} \frac{\epsilon}{2^{(k+1)+j+1}}
\end{equation}
holds. Hence, by equations \ref{n1_a}-\ref{n2_a}, the inequality
\begin{equation}\label{eq_trick}
\displaystyle   \int_{\mathcal{X}^{\leq n} }  \!\!\Phi^{\mu X.G_{1}}_{\rho} \, \, d \mathbb{P}^{\langle p, G_{1}\rangle}_{\sigma_{1},\sigma^{[k]}_{2}}\leq\int_{\mathcal{X}^{0} } \!\! \Phi^{\mu X.G_{1}}_{\rho} \, \, d \mathbb{P}^{\langle p, G_{1}\rangle}_{\sigma_{1},\sigma^{[k]}_{2}} + \Big( \displaystyle  \sum_{i\in I} \pi(\vec{t}_{i}) \cdot\displaystyle \big( \sem{G_{1}}_{\rho_{\gamma}}(q_{i}) \big)\Big)+ \displaystyle \sum_{j<n} \frac{\epsilon}{2^{(k+1)+j+1}}
\end{equation}
holds.

Let us now consider the Markov chain (depicted  in Figure \ref{fig4_prime}) in the game $\mathcal{G}^{G_{1}}_{\rho_{\gamma}}$, obtained from $M^{\langle p,G_{1}\rangle}_{\sigma_{1},\sigma_{2}}$ by removing the sub-Markov chains $M_{i}$, for $i\!\in\! I$. It follows  from definition of $\sigma^{[k]}_{2}$  that this is precisely the Markov chain induced by the strategy profile $\langle \tau_{1},\tau^{\frac{\epsilon}{2^{k+1}}}_{2}\rangle$, where $\tau_{1}$ is the strategy for Player $1$ in the game $\mathcal{G}^{G_{1}}_{\rho_{\gamma}}$ which behaves as the strategy $\sigma_{1}$ in the game $\mathcal{G}^{G_{1}}_{\rho_{\gamma}}$ until a terminal state of the form $\langle q,X\rangle$ is reached.
\begin{figure}
\centering

\pstree[ treemode=U,levelsep=8ex ]{\Tr{$\langle p, G_{1}\rangle$}}{
        \pstree[linestyle=none,arrows=-,levelsep=2ex]{\Tfan[fansize=10ex]}{\TR{ $ \mathcal{X}^{0} $}}
\pstree[levelsep=5ex, linestyle=none]{\Tr{$\langle q_{0},X\rangle$} \tlput{$\vec{t}_{0}$}\trput{$...$}}{
\Tr{$$}	
}
\pstree[levelsep=5ex, linestyle=none]{\Tr{$\ \ \ \ \langle q_{i},X\rangle\ \dots $} \trput{$\vec{t}_{i}\ ...$}}{
\Tr{$$}
}
}
\caption{Markov chain $M^{\langle p, G_{1}\rangle}_{\tau_{1},\tau^{\frac{\epsilon}{2^{k+1}}}_{2}}$ in  $\mathcal{G}^{G_{1}}_{\rho_{\gamma}}$}
 \label{fig4_prime}
\end{figure}  
The following equations are easy to verify:
\begin{center}
\begin{tabular}{l l l }
$E(M^{\langle p, G_{1}\rangle}_{\tau_{1},\tau^{\frac{\epsilon}{2^{k+1}}}_{2}})$ & $\bydef$ &  $\displaystyle \int_{\mathcal{P}} \Phi^{ G_{1}}_{\rho_{\gamma}}\ d \mathbb{P}^{\langle p, G_{1}\rangle}_{\tau_{1},\tau^{\frac{\epsilon}{2^{k+1}}}_{2}}$\\
$$ & $=_{A}$ & $\displaystyle \int_{\mathcal{X}^{0}} \Phi^{G_{1}}_{\rho_{\gamma}}\ d \mathbb{P}^{\langle p, G_{1}\rangle}_{\tau_{1},\tau^{\frac{\epsilon}{2^{k+1}}}_{2}} + \sum_{i\in I} \pi(\vec{t}_{i})\cdot \Phi^{G_{1}}_{\rho_{\gamma}} (\vec{t}_{i})$ \\
$$ & $=_{B}$ & $\displaystyle \int_{\mathcal{X}^{0}} \Phi^{G_{1}}_{\rho_{\gamma}}\ d \mathbb{P}^{\langle p, G_{1}\rangle}_{\tau_{1},\tau^{\frac{\epsilon}{2^{k+1}}}_{2}} + \sum_{i\in I} \pi(\vec{t}_{i})\cdot \rho_{\gamma}(X)(q_{i})$\\
$$ & $=_{C}$ & $\displaystyle \int_{\mathcal{X}^{0}} \Phi^{\mu X.G_{1}}_{\rho}\ d \mathbb{P}^{\langle p, G_{1}\rangle}_{\tau_{1},\tau^{\frac{\epsilon}{2^{k+1}}}_{2}} + \sum_{i\in I} \pi(\vec{t}_{i})\cdot \rho_{\gamma}(X)(q_{i})$\\
$$ & $=_{D}$ & $\displaystyle \int_{\mathcal{X}^{0}} \Phi^{\mu X.G_{1}}_{\rho}\ d \mathbb{P}^{\langle p, G_{1}\rangle}_{\tau_{1},\tau^{\frac{\epsilon}{2^{k+1}}}_{2}} + \sum_{i\in I} \pi(\vec{t}_{i})\cdot \sem{G_{1}}_{\rho_{\gamma}}(q_{i})$. 
\end{tabular}
\end{center}
Step $A$ is justified by the fact that every path in the set $\mathcal{X}^{0}$ defined earlier is also a path in the game $\mathcal{G}^{G_{1}}_{\rho_{\gamma}}$. Step $B$ follows from the fact that, by definition, $\Phi^{G_{1}}_{\rho_{\gamma}}(\vec{t}_{i})= \rho_{\gamma}(X)(q_{i})$, for every $i\!\in\! I$ and $q_{i}\!=\!last(\vec{t}_{i})$. Equation $C$ follows from Equation \ref{fix_step_aux_4}. Lastly, Equation $D$ follows from definition of $\rho_{\gamma}(X)$.

 By definition, the strategy $\tau^{\frac{\epsilon}{2^{k+1}}}_{2}$ is $\frac{\epsilon}{2^{k+1}}$-optimal (see Inequality \ref{delta_optimal_strat_1}). Thus, it follows from Equation \ref{eq_trick}  that the inequality 
\begin{equation}
\displaystyle   \int_{\mathcal{X}^{\leq n} }  \!\!\Phi^{\mu X.G_{1}}_{\rho} \, \, d \mathbb{P}^{\langle p, G_{1}\rangle}_{\sigma_{1},\sigma^{[k]}_{2}} \leq (\sem{G_{1}}_{\rho_{\gamma}}(p) +  \frac{\epsilon}{2^{k+1}} )+ \displaystyle \sum_{j<n} \frac{\epsilon}{2^{(k+1)+j+1}}
\end{equation}
or equivalently, 
\begin{center}
$\displaystyle   \int_{\mathcal{X}^{\leq n} }  \!\!\Phi^{\mu X.G_{1}}_{\rho} \, \, d \mathbb{P}^{\langle p, G_{1}\rangle}_{\sigma_{1},\sigma^{[k]}_{2}} \leq \sem{G_{1}}_{\rho_{\gamma}}(p) +   \displaystyle \sum_{j\leq n} \frac{\epsilon}{2^{k+j+1}}$
\end{center}
holds. We have then proved that Equation \ref{strategy_eq_3} holds, as desired. Therefore, following backwards our previous analysis, Equation \ref{strategy_eq_2} and, thus,  Equation \ref{fix_step_aux_3_prime} hold, and this concludes the proof.

\textbf{Inductive case} $\mathbf{G\!=\! \nu X.G_{1}}$. \\
Similar to the previous one.

\section{Conclusions and future work}
We proved that the denotational and game semantics of \cite{MM07} of the logic pL$\mu$ coincide on all PLTS's. This result, which is yet another example of  application of game theory to logic, strengthen the theory of the logic pL$\mu$, which is recently emerging as an interesting tool for expressing properties and reasoning about PLTS's.

Further recent research \cite{MIO11, MioThesis}, explores the extension of the logic obtained by adding two new conjunction/disjunction operators called \emph{product} ($\cdot$) and \emph{coproduct} ($\odot$). The product operator, whose denotational semantics is defined as $\sem{F\cdot G}(p)\!=\!\sem{F}(p)\cdot \sem{G}(p)$, and the coproduct operator (the De Morgan dual of the product respect to the involution $\neg x \!=\! 1-x$) increases the expressive power of the logic. For instance it is possible to encode the \emph{qualitative} modality $\mathbb{P}_{>0}F$ whose semantics can be defined as $\sem{\mathbb{P}_{>0}F}(p)\!=\!1$ if $\sem{F}(p)\!>\!0$;  $\sem{\mathbb{P}_{>0}F}(p)\!=\!0$ otherwise. This allows the expression of interesting properties, as well as the encoding of important temporal probabilistic logics such as (qualitative) PCTL.

\section*{Acknowledgement}
The author would like to thank his PhD supervisor Alex Simpson and two anonymous referees for helpful  suggestions.  This research was supported by a full PhD scholarship provided by LFCS-School of Informatics at the University of Edinburgh, and by  the EPSRC research grant  ``Linear Observations and Computational Effects''.

\bibliographystyle{abbrv}
\bibliography{FICS2010journal}

\end{document}